\documentclass[twocolumn,times,twocolappendix]{aastex63}
\usepackage{amsmath}
\usepackage{color}

\shorttitle{WA in Radiation-driven Fountain}
\shortauthors{Ogawa et al.}

\begin{document}

\title{Warm Absorbers in the Radiation-driven Fountain Model of
Low-mass Active Galactic Nuclei}

\correspondingauthor{Shoji Ogawa}
\email{ogawa@kusastro.kyoto-u.ac.jp}

\author[0000-0002-5701-0811]{Shoji Ogawa}
\affil{Department of Astronomy, Kyoto University, Kitashirakawa-Oiwake-cho, Sakyo-ku, Kyoto 606-8502, Japan}

\author[0000-0001-7821-6715]{Yoshihiro Ueda}
\affiliation{Department of Astronomy, Kyoto University, Kitashirakawa-Oiwake-cho, Sakyo-ku, Kyoto 606-8502, Japan}

\author[0000-0002-8779-8486]{Keiichi Wada}
\affiliation{Kagoshima University, Graduate School of Science and Engineering, Kagoshima 890-0065, Japan}
\affiliation{Ehime University, Research Center for Space and Cosmic Evolution, Matsuyama 790-8577, Japan}
\affiliation{Hokkaido University, Faculty of Science, Sapporo 060-0810, Japan}

\author[0000-0003-2161-0361]{Misaki Mizumoto}
\affiliation{Department of Astronomy, Kyoto University, Kitashirakawa-Oiwake-cho, Sakyo-ku, Kyoto 606-8502, Japan}
\affiliation{Hakubi Center, Kyoto University, Yoshida-honmachi, Sakyo-ku, Kyoto, 606-8501, Japan}

\begin{abstract}

To investigate the origins of the warm absorbers in active galactic
nuclei (AGNs), we study the ionization-state structure of the
radiation-driven fountain model in a low-mass AGN \citep{Wada2016} 
and calculate the predicted X-ray spectra, utilizing
the spectral synthesis code \textsf{Cloudy} \citep{Ferland2017}. 
The spectra show many absorption and emission line features originated
in the outflowing ionized gas. The \ion{O}{8} 0.654~keV lines are produced
mainly in the polar region much closer to the SMBH than the optical
narrow line regions. The absorption measure distribution of the
ionization parameter ($\xi$) at a low inclination
spreads over 4 orders of magnitude in $\xi$,
indicating multi-phase ionization structure of the outflow, as
actually observed in many type-1 AGNs.
We compare our simulated spectra with the high
energy-resolution spectrum of the narrow line Seyfert 1 galaxy, NGC~4051.  
The model reproduces
slowly outflowing (a few hundreds km~s$^{-1}$) warm absorbers.
However, the faster components with a few thousands km~s$^{-1}$ observed in NGC~4051 are not reproduced.
The simulation also underproduces the intensity and width of the \ion{O}{8}
0.654~keV line.
These results suggest that the ionized gas launched from sub-parsec or smaller regions
inside the torus, which are not included in the current
fountain model, must be important ingredients of the warm absorbers with a few thousands km~s$^{-1}$.
The model also consistently explains the Chandra/HETG spectrum of the Seyfert 2 galaxy, the Circinus galaxy.

\end{abstract}


\section{Introduction}
\label{sec1}

Outflows in various physical conditions
are ubiquitously observed from an active
galactic nuclei (AGN). They constitute essential elements of the AGN
structure, such as jets, warm absorbers, narrow emission line regions
(NLRs), and tori \citep[e.g.,][]{Elitzur2006,Netzer2015,Wada2018b,Alonso-Herrero2021}. By carrying a
huge amount of mass, momentum, and energy from the nucleus, these
AGN-driven outflows play key roles in determining the dynamics of
accretion flow onto the supermassive black hole (SMBH), and even
significantly affect its environment in the host galaxy or larger
scale \citep[e.g.,][]{Fabian2012,Harrison2017,Veilleux2020}. Thus,
revealing the physical properties and origins of the outflows is
important to understand AGN feeding/feedback mechanisms.

Outflows of mildly ionized gas can be recognized in the ultraviolet (UV) to soft
X-ray ($\lesssim$2~keV) spectrum of an AGN by blue-shifted absorption
line and edge features \citep[see
  e.g.,][]{Kaastra2000,Kaastra2002}. They are phenomenologically
referred to as ``warm absorbers'', which are detected in
the X-ray spectra of about half of nearby type-1 AGNs
\citep[e.g.,][]{Reynolds1997,Laha2014}. With spectral fitting,
one can derive the ionization parameter of the absorbers,
$\xi = L_\mathrm{ion}/n_\mathrm{H}r^2$ (where $L_\mathrm{ion}$ is the
AGN luminosity\footnote{In this paper we define $L_\mathrm{ion}$ as an luminosity
integrated from 13.6~eV to 13.6~keV.}, $n_\mathrm{H}$ the hydrogen
number density, and $r$ the distance from the ionizing source), the
column density ($N_\mathrm{H}$), and the outflow velocity
($v_\mathrm{out}$).  It is also possible to constrain $n_\mathrm{H}$
and/or $r$ using other diagnostics, such as population ratios among
different energy levels of the same ions \citep[e.g.,][]{Mao2017} and time
variability \citep[e.g.,][]{Krongold2007}.  The typical outflow velocities of the warm absorbers
are a few hundred to a few thousand km~s$^{-1}$. They show wide
ranges of values in physical parameters such as $\xi$ and $N_\mathrm{H}$
\citep[e.g.,][]{Kaastra2002,Kaspi2004,Behar2017}, indicating
multiphase nature of the AGN-driven outflows. Since their outflow velocities are
low, the energy and momentum outflow rates carried by the warm absorbers
are smaller and hence have less impacts on the environments compared
with ultrafast outflows (UFOs), which have velocities of $\sim$0.1$c$
\citep[e.g.,][]{Tombesi2010}. However, due to their large mass outflow rates, the warm absorbers
are important to understand the global processes of mass flow in
AGN systems, e.g., what fraction of mass fed by the host galaxy is
eventually accreted by the SMBH and is ejected back into the
surroundings.

The physical origins of the warm absorbers still remain unclear, although
several theoretical models have been proposed \citep[e.g.,][]{Krolik1995,Proga2000,Fukumura2010}. Assuming
that the outflow velocities of the warm absorbers correspond to the escape
velocities from the gravitational potential of the SMBH, they are
likely to be launched from outer accretion disks and/or torus regions.
\citet{Mizumoto2019} have shown that the warm absorbers can be
explained as thermally-driven winds from
the broad line region and torus 
(see e.g., \citealt{Krolik1995} for earlier works). However, they assume a very
simplified geometry, whereas 
the real structures of the interstellar medium in the central 
region of AGNs are still unknown.
Hence, it is quite important to
investigate the origins of the warm absorbers on the basis of more
realistic, physically motivated dynamical models.

\cite{Wada2012} proposed a ``radiation-driven fountain'' model of an
AGN based on three-dimensional radiation–hydrodynamic simulations. In
this model, the outflows are driven mainly by the radiation pressure
to dust and the thermal energy of the gas acquired via X-ray heating,
and a geometrically thick torus-like shape is naturally formed (see
Figure~\ref{fig_wada_model}). \citet{Wada2016} applied this
radiation-driven fountain model to the Circinus galaxy, which is one
of the closest \citep[4.2~Mpc:][]{Freeman1977} type-2 AGN.  
The model is consistent with several observed features, such as the infrared spectral energy
distribution (SED, \citealp{Wada2016}), the dynamics of
atomic/molecular gas \citep{Wada2018a,Izumi2018,Uzuo2021},
distribution of ionized gas in the NLR \citep{Wada2018b}, and
broadband X-ray spectra \citep{Buchner2021}.
\citet{Mizumoto2019} also suggested that this kind of dust driven wind
may play an important role to the warm absorber acceleration.

In this paper, we present the results of mock X-ray high energy-resolution
spectroscopic observations based on the radiation-driven fountain in a low
mass AGN produced by \citet{Wada2016}.
We adopt the same approach as
in \citet{Wada2018b}, which calculated the intensity maps of optical
emission lines (H$\alpha$, H$\beta$, [\ion{O}{3}], [\ion{N}{2}], and [\ion{S}{2}])
utilizing the \textsf{Cloudy} code \citep{Ferland2017}. 
The three dimensional density and velocity maps in a snapshot of the
\citet{Wada2016} model are used as an input. 
Combining \textsf{Cloudy}
runs along the radial direction, we make the map of the ionization
state and calculate both the transmitted and
scattered X-ray spectra at
various inclination angles, which contain absorption and emission
lines, respectively.
Then, we compare these mock spectra with actual observations and 
investigate whether the observed spectral features of ionized absorbers
(including both absorption and emission lines)
can be explained by the radiation-driven
fountain model.

We choose NGC~4051 (narrow line Seyfert 1) as the comparison target.
It has similar AGN parameters to those assumed in
the \citet{Wada2016} model; the black hole mass, bolometric
luminosity, and Eddington ratio of NGC~4051 are ($\log
M_\mathrm{BH}/M_\odot$, $\log L_\mathrm{bol}$, $\lambda_\mathrm{Edd}$)
= ($5.6$, $43.2$, $0.33$)\footnote{The references for the black hole
  mass and Eddington ratio are \citet{Koss2017} and 
  \citet{Ogawa2021}, respectively.}.
Extensive archival data of NGC~4051 observed with XMM-Newton are available.
The high energy-resolution spectrum
observed with the
Reflection Grating Spectrometer (RGS) on XMM-Newton shows a number of
complex emission/absorption features indicative of multiple warm
absorbers.
Thus, NGC~4051 is an ideal target for our study.

The structure of this paper is as
follows. Section~\ref{sec2} describes the overview of the
radiation-driven fountain model and the numerical methods of our
\textsf{Cloudy} simulations. The results of the simulations are
summarized in Sections~\ref{sec3}. In Section~\ref{sec4}, we compare
our results with the observed X-ray spectrum of NGC~4051 
and discuss the implications.
Fe K$\alpha$ intensity distributions and 
comparison with the X-ray spectrum of the Circinus galaxy are shown in Appendix~\ref{A1}.

\section{Simulations}
\label{sec2}

\subsection{Input Model}
\label{sec2.1}

\begin{figure*}
\plottwo{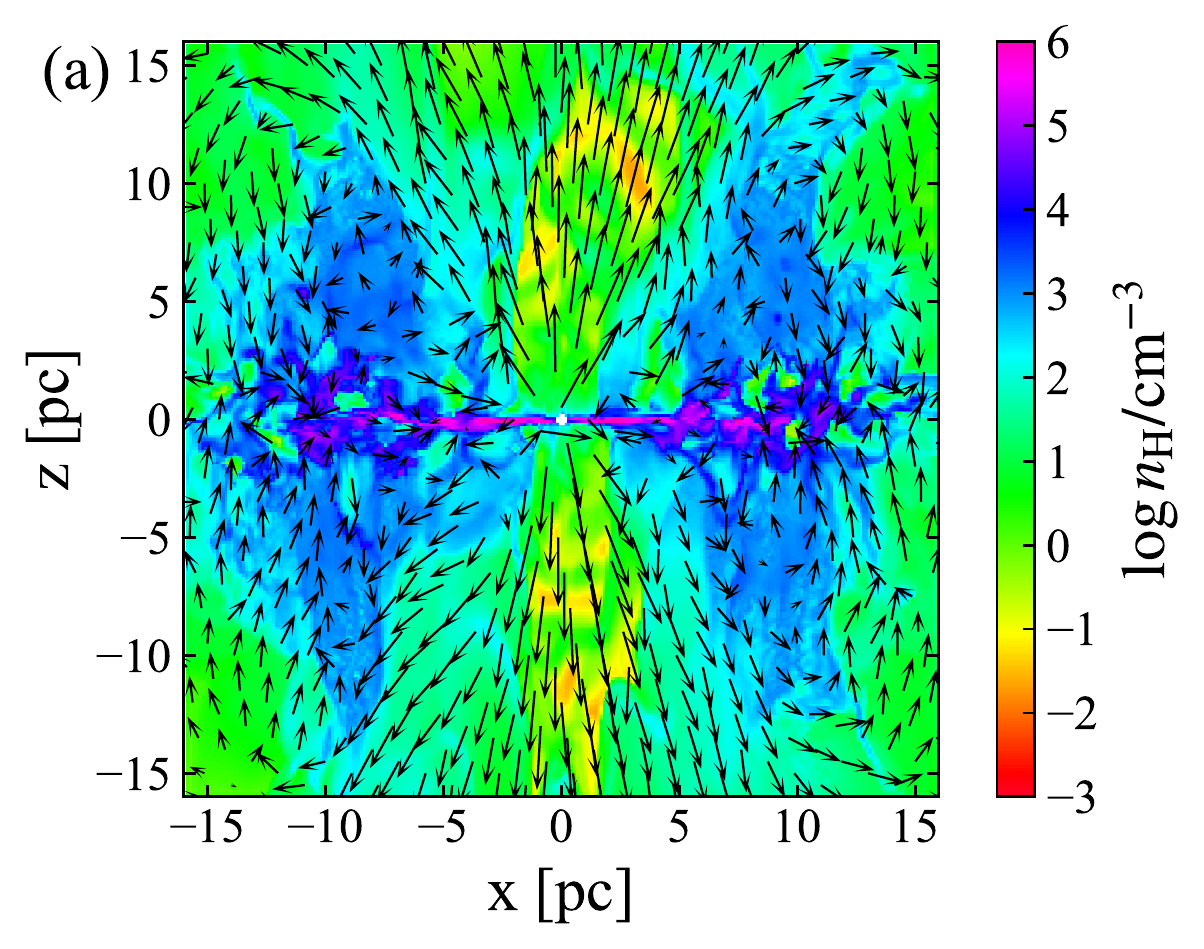}{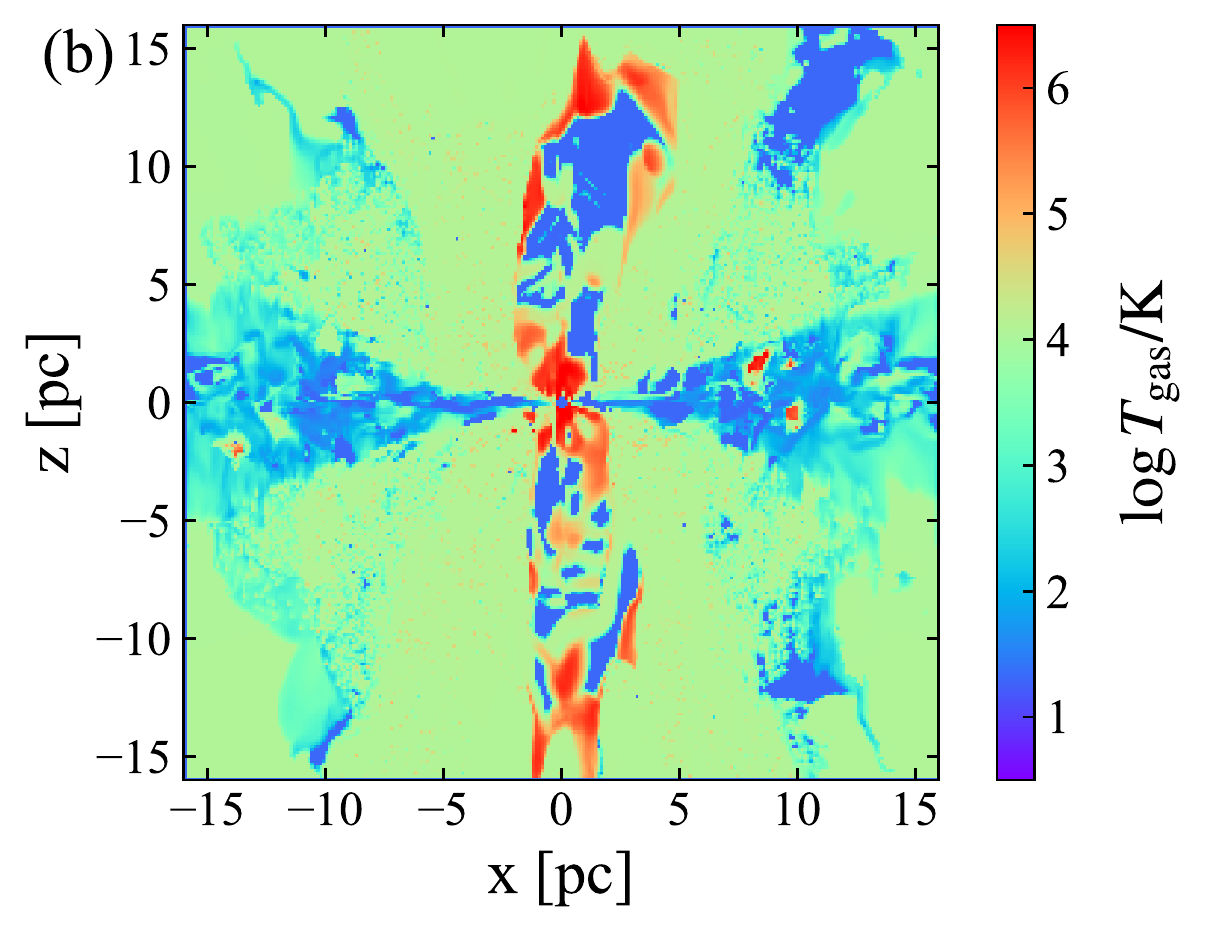}
\caption{
(a): A snapshot of hydrogen number-density distribution on a x-z plane
  in the radiation-driven fountain model \citep{Wada2016}.
  The arrows represent the relative velocity fields in logarithmic scales. 
(b): Same as (a), but for temperature distribution. 
}
\label{fig_wada_model}
\end{figure*}

\begin{figure*}
\epsscale{1.0}
\plottwo{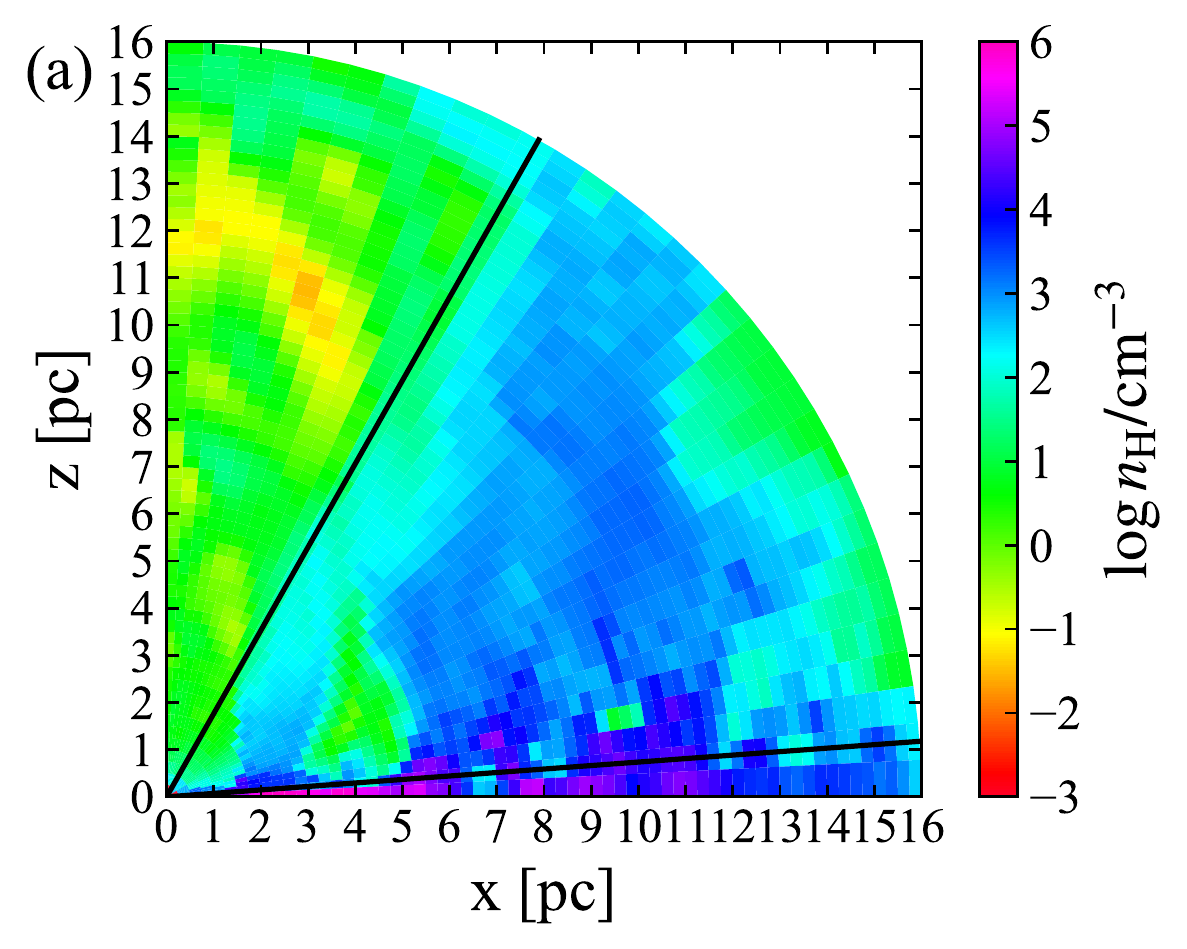}{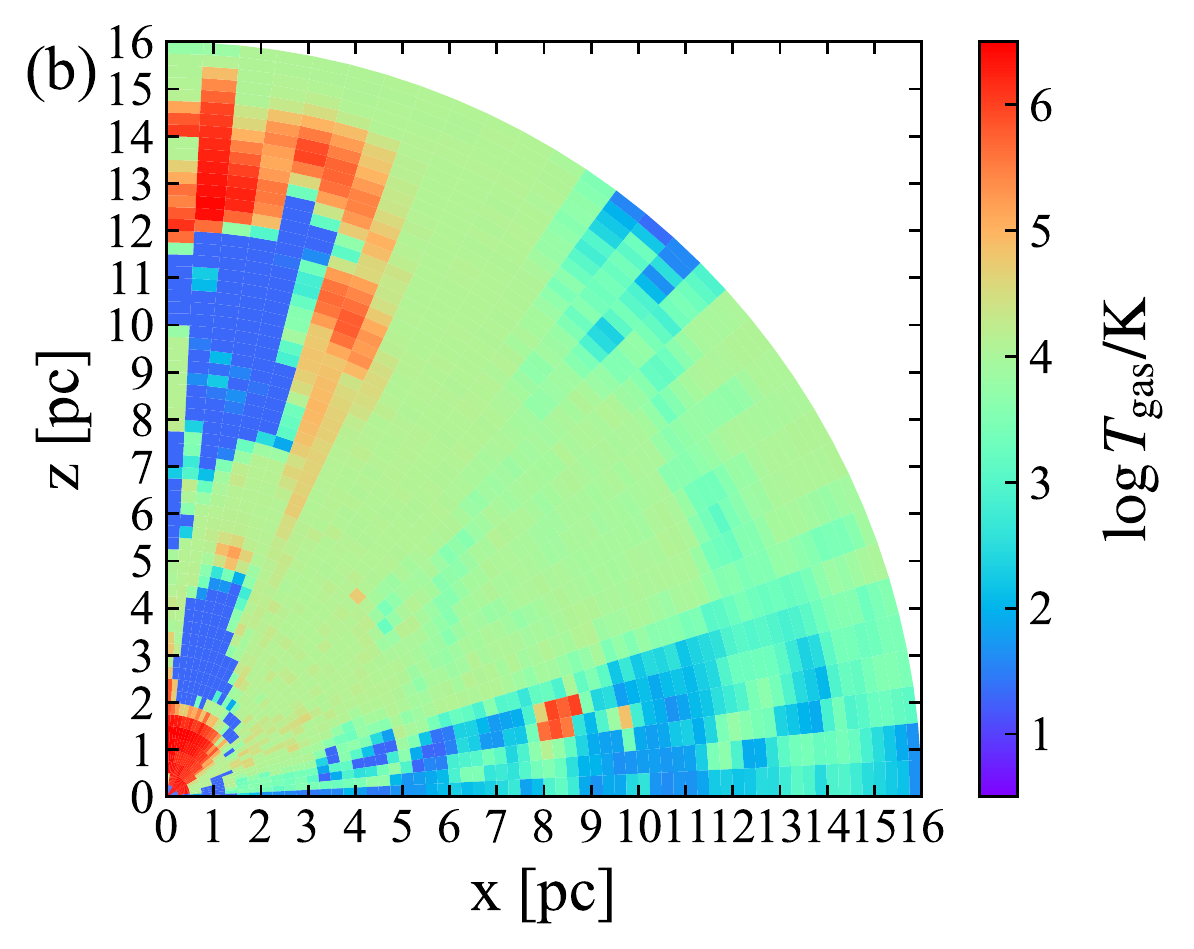}
\epsscale{0.58}
\plotone{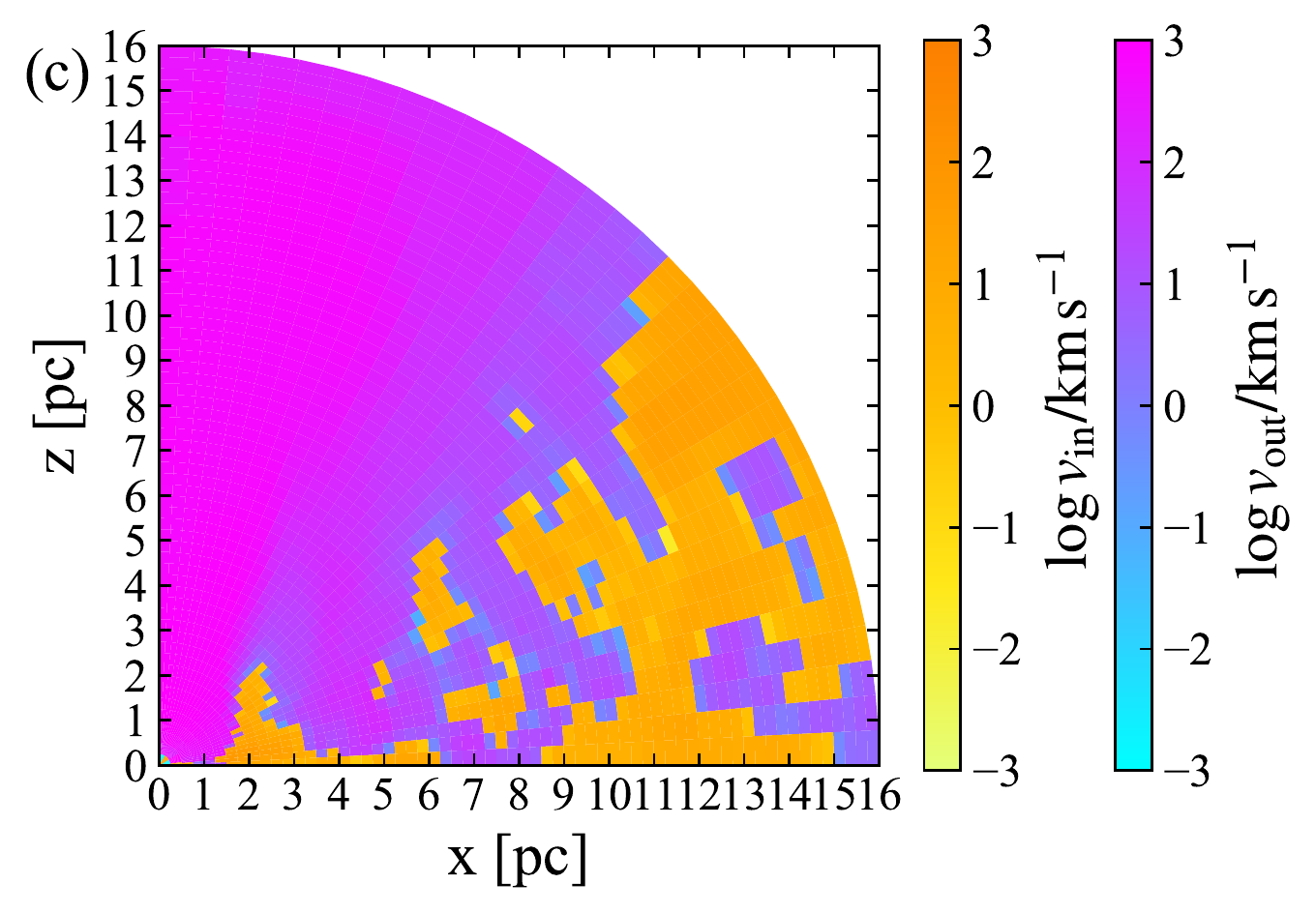}
\caption{
(a): Hydrogen number-density ($n_\mathrm{H}$) structure in a polar grid reformed from the original data in the radiation-driven fountain model. 
Black lines represents the lines of sight for $i=$ 30 and 86~degrees from z-axis.
(b): Same as (a), but for the temperature ($T_\mathrm{gas}$) structure.
(c): Structure of radial inflow or outflow velocities along the lines of sight ($v_\mathrm{in}$ and $v_\mathrm{out}$).
}
\label{fig_reformed}
\end{figure*}

Here we summarize the assumptions of the radiation-driven fountain
model by \citet{Wada2016}, one of the dynamical models of AGN tori
with supernova feedback based on three dimensional radiation-hydrodynamic
simulations.
The model only considers matter distribution in the torus region (0.125~pc $< r <$ 16~pc)
but accounts for
radiative feedback processes from the AGN such as the radiation
pressure on the surrounding material and X-ray heating.
The non-spherical radiation field caused by the AGN is considered.
They set the black hole mass of $M_\mathrm{BH}$ to $2 \times 10^6
M_\odot$,
the Eddington ratio $\lambda_\mathrm{Edd}$ to $0.2$, the bolometric
luminosity $L_\mathrm{bol}$ to $5 \times 10^{43}$~erg~s$^{-1}$, and the X-ray
luminosity (2--10~keV) $L_\mathrm{X}$ to $2.8 \times
10^{42}$~erg~s$^{-1}$.
The solar metallicity and cooling functions for 20~K $\leq T_\mathrm{gas} \leq$ $10^8$~K \citep{Meijerink2005,Wada2009} are assumed, where $T_\mathrm{gas}$ is the temperature of gas. 
The hydrodynamics calculations cover (32~pc)$^3$ region in the $256^3$
grid cells. Figure~\ref{fig_wada_model}(a) and (b) show the
distributions of gas density and temperature, respectively.  The
velocity field of gas is over-plotted in
Figure~\ref{fig_wada_model}(a).

\subsection{Radiative Transfer}

\begin{figure}
\plotone{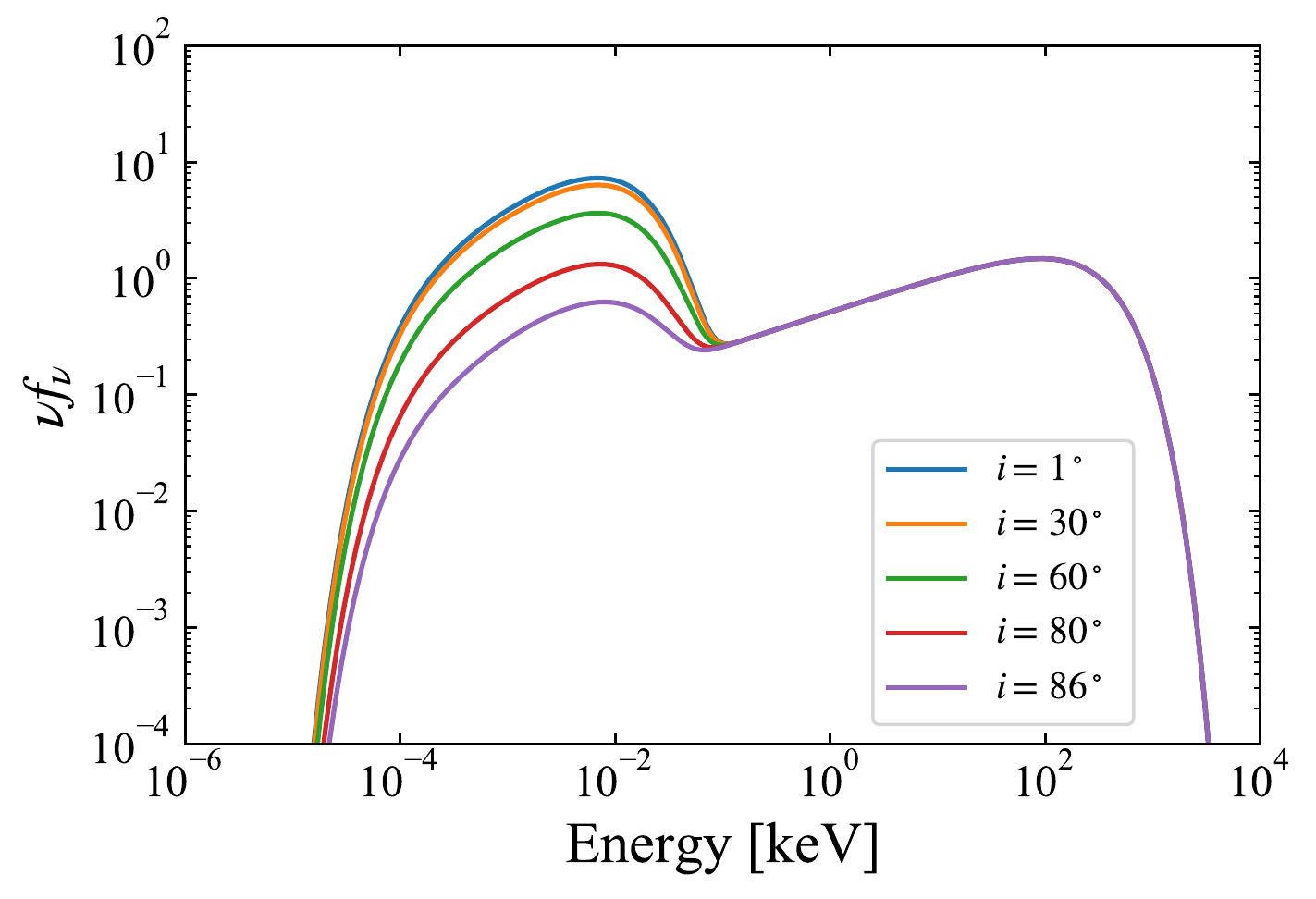}
\caption{
  Incident SED models of the AGN at the innermost grids for
  inclination angle of $i=$ 1, 30, 60, 80, and 86~degrees (from top to
  bottom).
The vertical axis has an arbitrary unit of $\nu f_\nu$, where $f_\nu$ is the energy flux at the frequency $\nu$.
}
\label{fig-insed}
\end{figure}

Following the quasi-three dimensional radiative transfer calculations
in \citet{Wada2018b}, we make use of \textsf{Cloudy}~v17.02 \citep{Ferland2017} to produce the ionization-state map and to simulate the spectra. 
We reform the cells from the $256^3$ Cartesian grid to the $64^3$ uniformly
spaced polar grid.
Figure~\ref{fig_reformed} plots the geometry of the reformed model.
We input the 3-dimensional maps of physical 
parameters (hydrogen density and temperature), obtained 
in a snapshot of the radiation-driven fountain model, into the
\textsf{Cloudy} code. Figure~\ref{fig-insed} shows the input SED
model at the innermost grids. It is equivalent to that 
obtained by the \textsf{Cloudy}'s \textsf{AGN} command and is represented as:
\begin{eqnarray}
\label{eqn:agncon}
f \left(\nu\right)    =  \nu ^{\alpha _\mathrm{UV} } \exp \left( { - h\nu /kT_\mathrm{BB} } \right)\exp \left( { - kT_\mathrm{IR} /h\nu } \right)\cos{i} \nonumber\\
                         +  a\nu ^{\alpha_\mathrm{X} } \exp \left( { - h\nu /E_1 } \right) \exp \left( { - E_2 /h\nu } \right),
\end{eqnarray}
where 
$\alpha _\mathrm{UV} = -0.5$, $T_\mathrm{BB} = 10^5$~K, 
$\alpha_\mathrm{X} = -0.7$, $a$ is a constant that yields the X-ray to UV ratio $\alpha_\mathrm{OX} = -1.4$, $kT_\mathrm{IR} = 0.01$~Ryd,
$E_\mathrm{1} = 300$~keV, $E_\mathrm{2} = 0.1$~Ryd, and $i$ is the angle from the z-axis (i.e., inclination). 
The UV radiation (first term), which comes from the geometrically
thin, optically-thick disk, 
is assumed to be proportional to $\cos{i}$. 
Whereas, the X-ray component (second term) is assumed to be isotropic.

We run a sequence of \textsf{Cloudy} simulations along
the radial direction from the central source to the outer boundary at $r=16$~pc,
where the output spectrum
(net transmitted spectrum, which is the sum of the attenuated incident radiation and the diffuse emission emitted by the photo-ionized plasma)
from a cell is used as the input spectrum to the next (outer) cell by
taking into account the velocity field of each cell.
redHere, we assume that the density structure in each cell is uniform.\footnote{This assumption should be 
verified by future numerical simulations with a higher resolution,
but we confirmed that the spectral properties of the narrow line region 
which is also originated in the outflow are reproduced with the same assumption \citep[see][]{Wada2018b}.} 
For simplicity, we also assume that the net transmitted radiation
is the primary 
ionizing source and 
the scattered radiation from the surrounding regions out of the line of sight
is ignored.
To confirm the validity of this assumption, we calculate the
differences of the ionization parameters $\xi$ when the
scattered components
from the adjacent cells out of the line of sight
are added to the input spectrum in each run.
To estimate the scattered components, we utilize
``the reflected spectra'' calculated by \textsf{Cloudy}, which
contains Compton-scattered incident radiation and the diffuse emission.
We find that 
the increase in the $\xi$ parameter is $\sim$2\% on average, 
except for a few grid cells in
the equatorial plane ($>$100\%). 
Thus, we restrict the analysis for the inclination angle to $i \leq 88^\circ$. 
After we calculate the ionization state and spectra in all cells, we
integrate the reflected spectra
over all cells by taking into
account the optical depths in each cell
along the line of sight. We refer to it as ``scattered spectra'' in
the following.

\section{Results}
\label{sec3}

\subsection{Ionization Structure}

\begin{figure}
\plotone{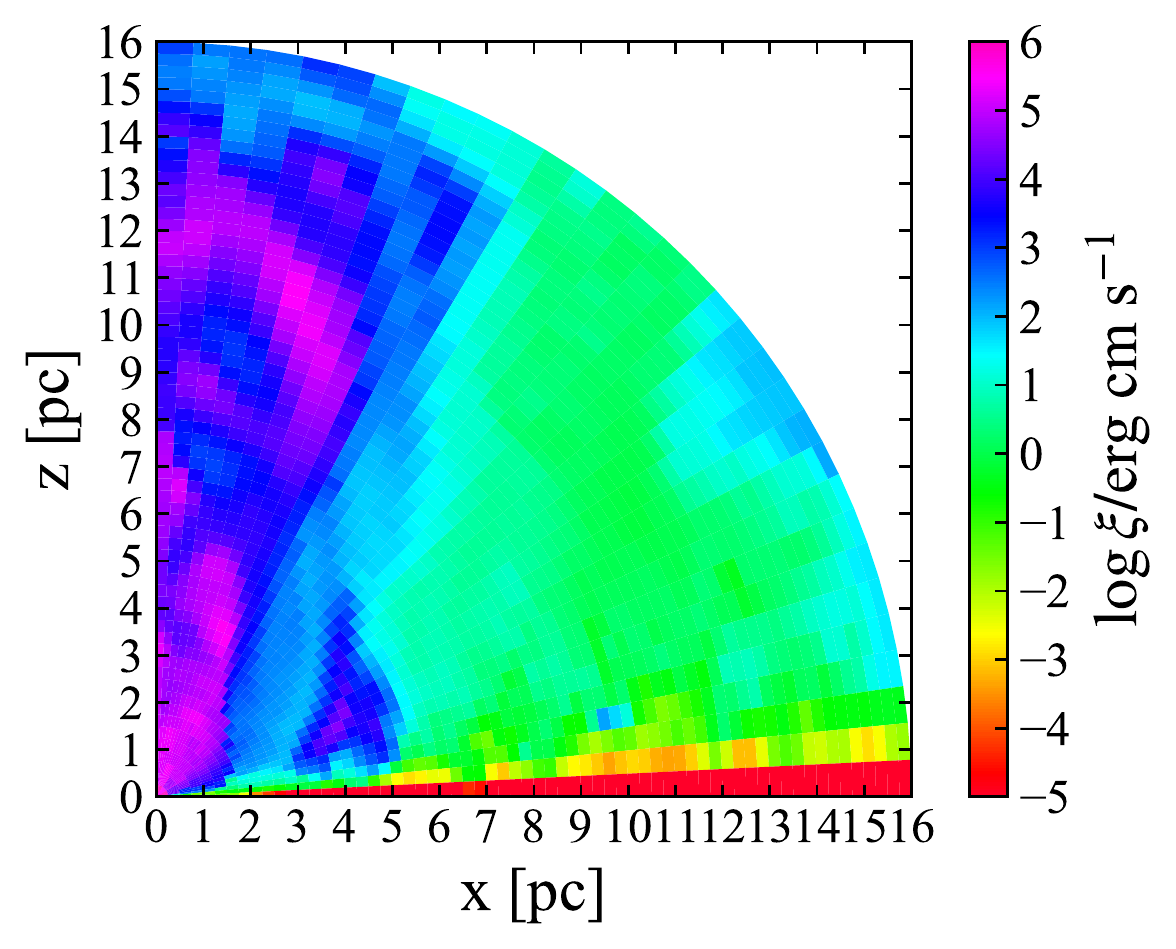}
\caption{
  Structure of the ionization parameter ($\xi$) obtained
  by the \textsf{Cloudy} calculation.
}
\label{fig-ip}
\end{figure}

\begin{figure}
\plotone{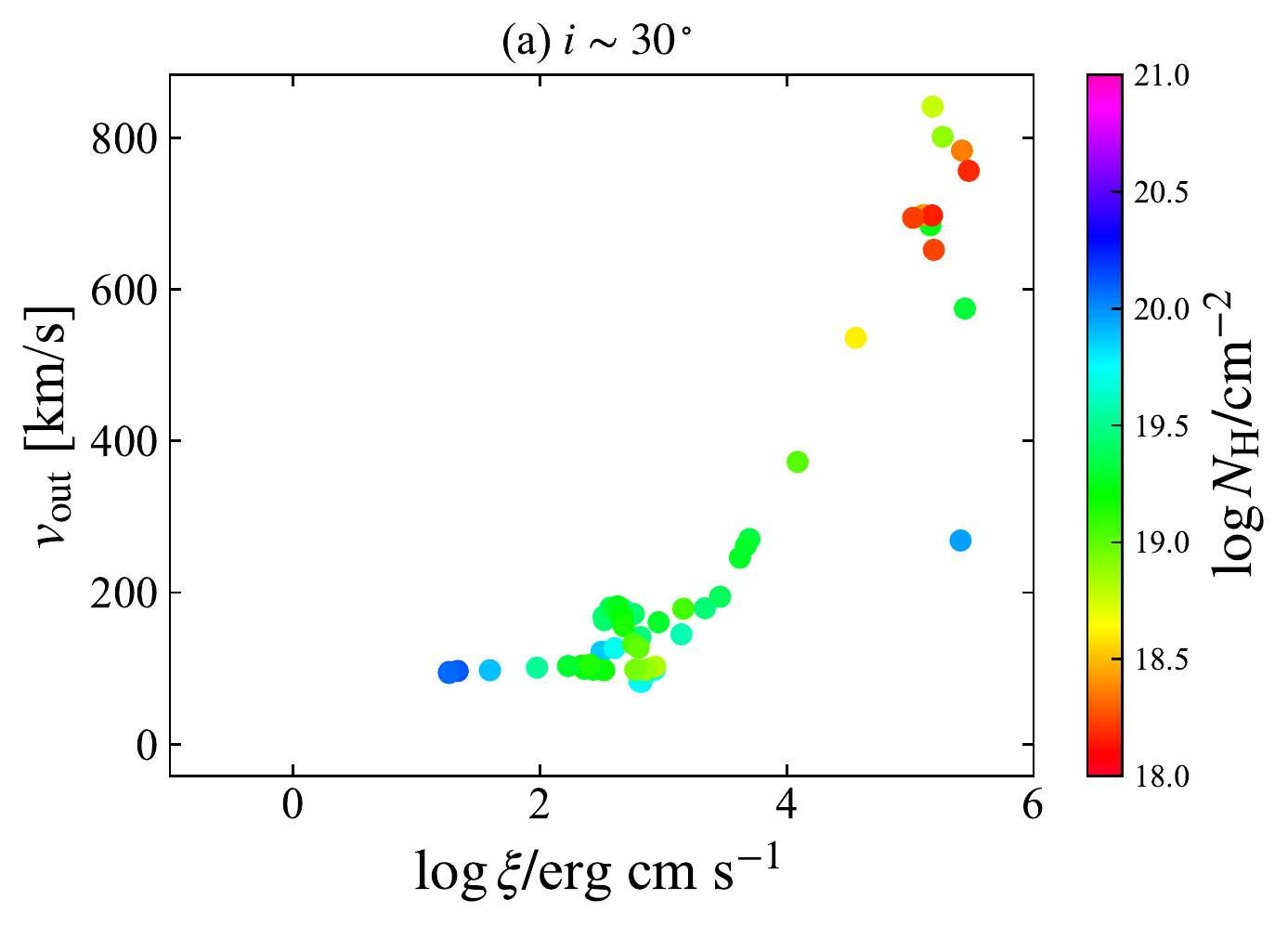}
\plotone{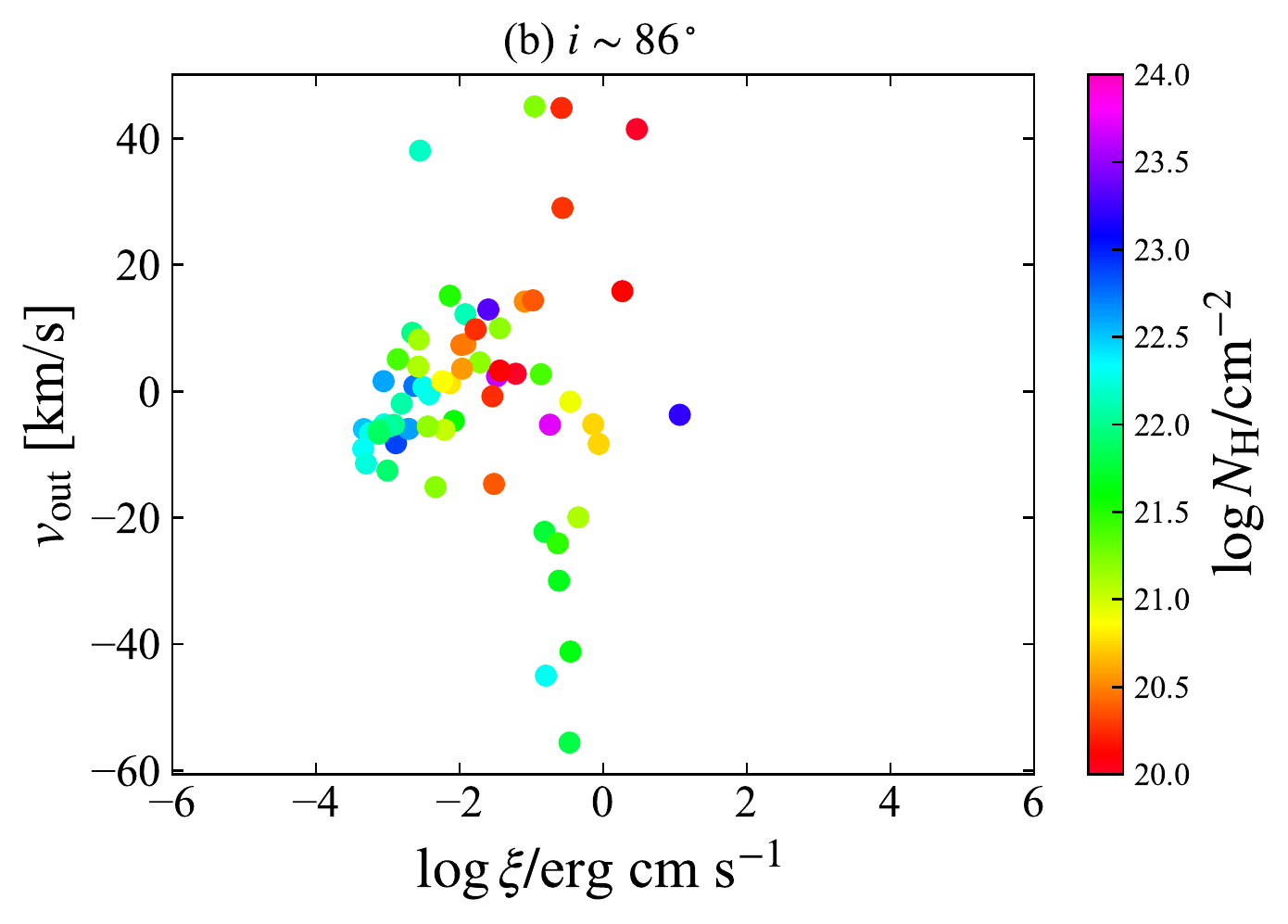}
\caption{
Relation between the ionization parameter ($\xi$) and the
outflow velocity ($v_\mathrm{out}$) for the grid cells along a line of sight. 
Each data point is color-coded by the hydrogen column density in the cell.
(a): for $i = 30$~degrees. 
(b): for $i = 86$~degrees.
}
\label{fig-vxi}
\end{figure}

Figure~\ref{fig-ip} plots the map of the ionization parameter, $\xi =
L_\mathrm{ion}/n_\mathrm{H} r^2$.
The ionization parameter is distributed over a wide range
and shows a complex structure, which 
mainly reflects that of hydrogen density
(Figure~\ref{fig_reformed}(a)).
It is also seen that non-isotropic radiation from the central
accretion disk makes the gas in the polar region more highly ionized ($\log \xi > 3$)
than that in higher inclination regions with the same hydrogen density
and distance.

Figure~\ref{fig-vxi} plots the relation between
the ionization
parameter ($\xi$) and the outflow velocity ($v_\mathrm{out}$) for the
cells along a line of sight, where each data point is color-coded
by the hydrogen column density of that cell.
We show the results for two inclination angles, $i=$ 30 and 86~degrees,
which are representative of type-1 and type-2 AGNs, respectively
(see section~\ref{sec4.2} and Appendix~\ref{A1}).
As noticed, 
at the low inclination angle (Figure~\ref{fig-vxi}(a)), 
$\xi$ is well correlated with $v_\mathrm{out}$, except for the highest
ionization states corresponding to the closest region to the SMBH.
In the polar region, the gas is almost spherically expanding,
and therefore from mass conservation, we expect that
$v_\mathrm{out} \propto 1/(n_\mathrm{H} r^2)$ 
and hence $\xi \propto v_\mathrm{out}$. 
By contrast, no clear correlation is found for the high inclination case
(Figure~\ref{fig-vxi}(b)).
In the region closer to the equatorial plane, the gas motion is dominated by random motion caused by the backflow, turbulence, and supernova feedback, rather than the coherent motion of outflow. This makes $v_\mathrm{out}$ smaller ($<$40--60~km~s$^{-1}$) and more complex.
See for example, \citet{Wada2002,Kawakatu2008},  
on the effect of supernova feedback on the turbulent structures of the circumnuclear disk. 
Note that the radiation-driven outflows from which the warm absobers are originated are not directly affected by the supernova feedback.
It is notable that the column density is large ($\log N_\mathrm{H} > 24$ in total) and radial velocity is relatively slow (a few tens of km~s$^{-1}$) in this region.

\subsection{X-ray Spectra}

\begin{figure*}
\epsscale{1.0}
\plottwo{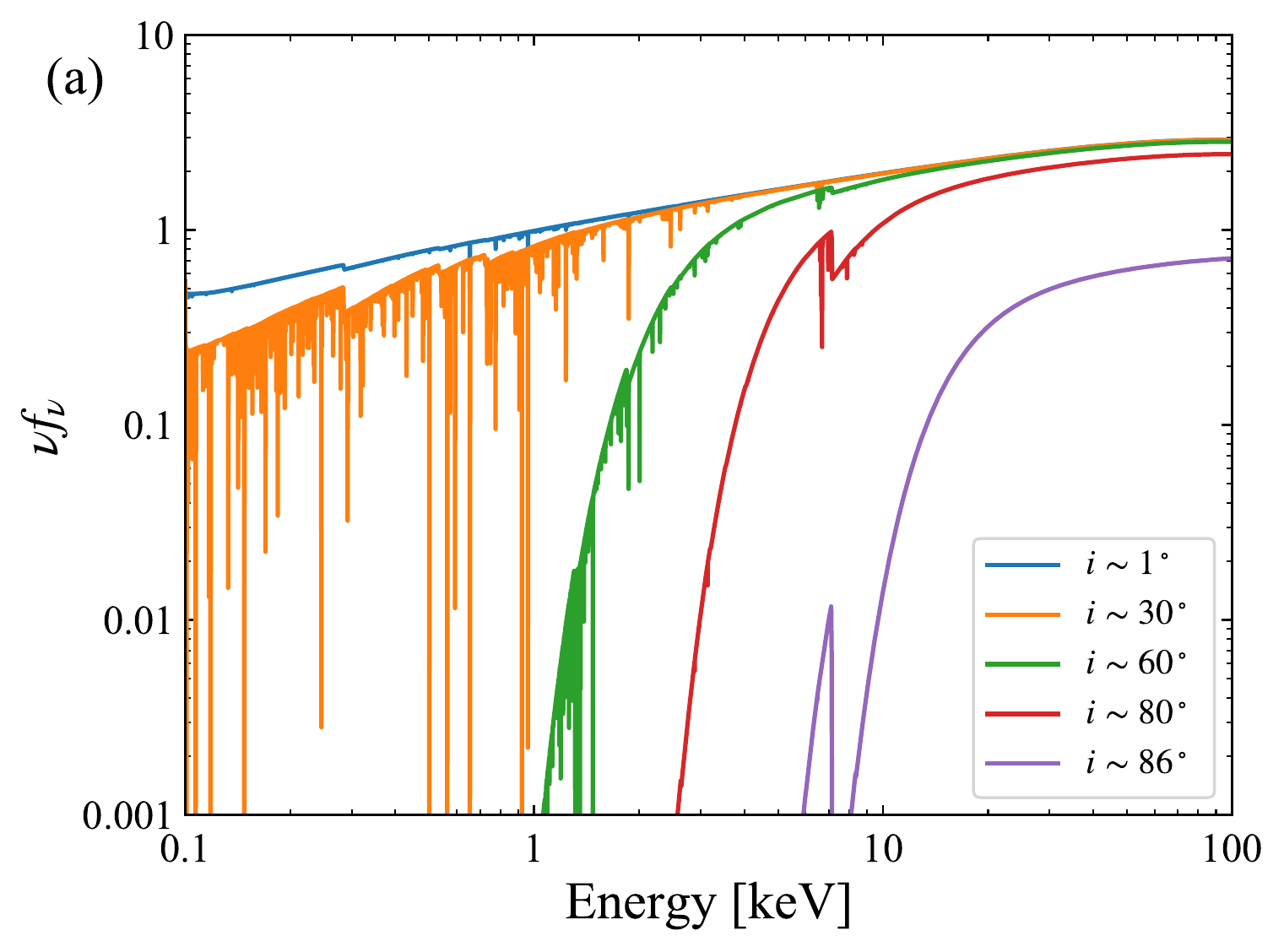}{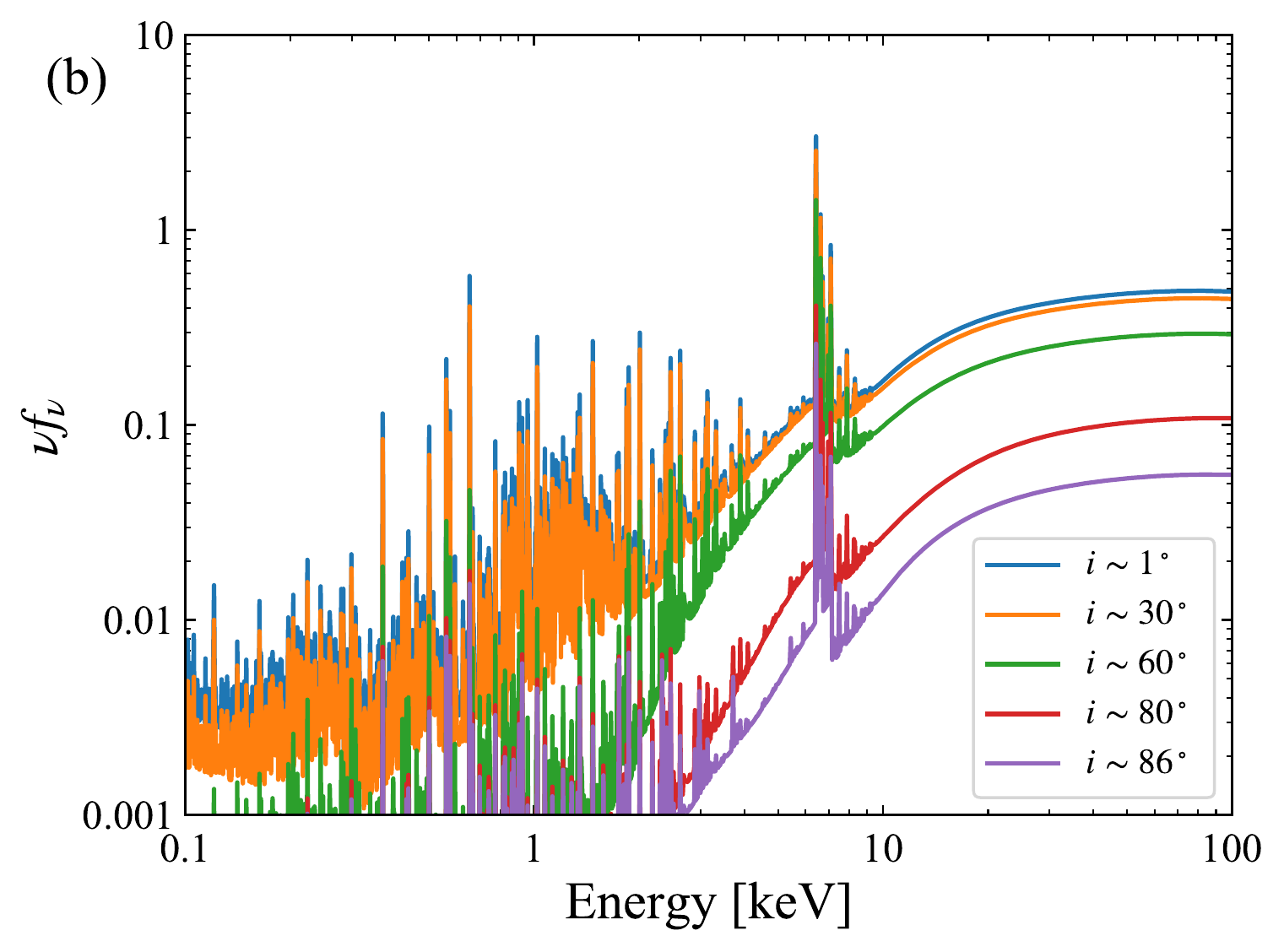}
\epsscale{0.5}
\plotone{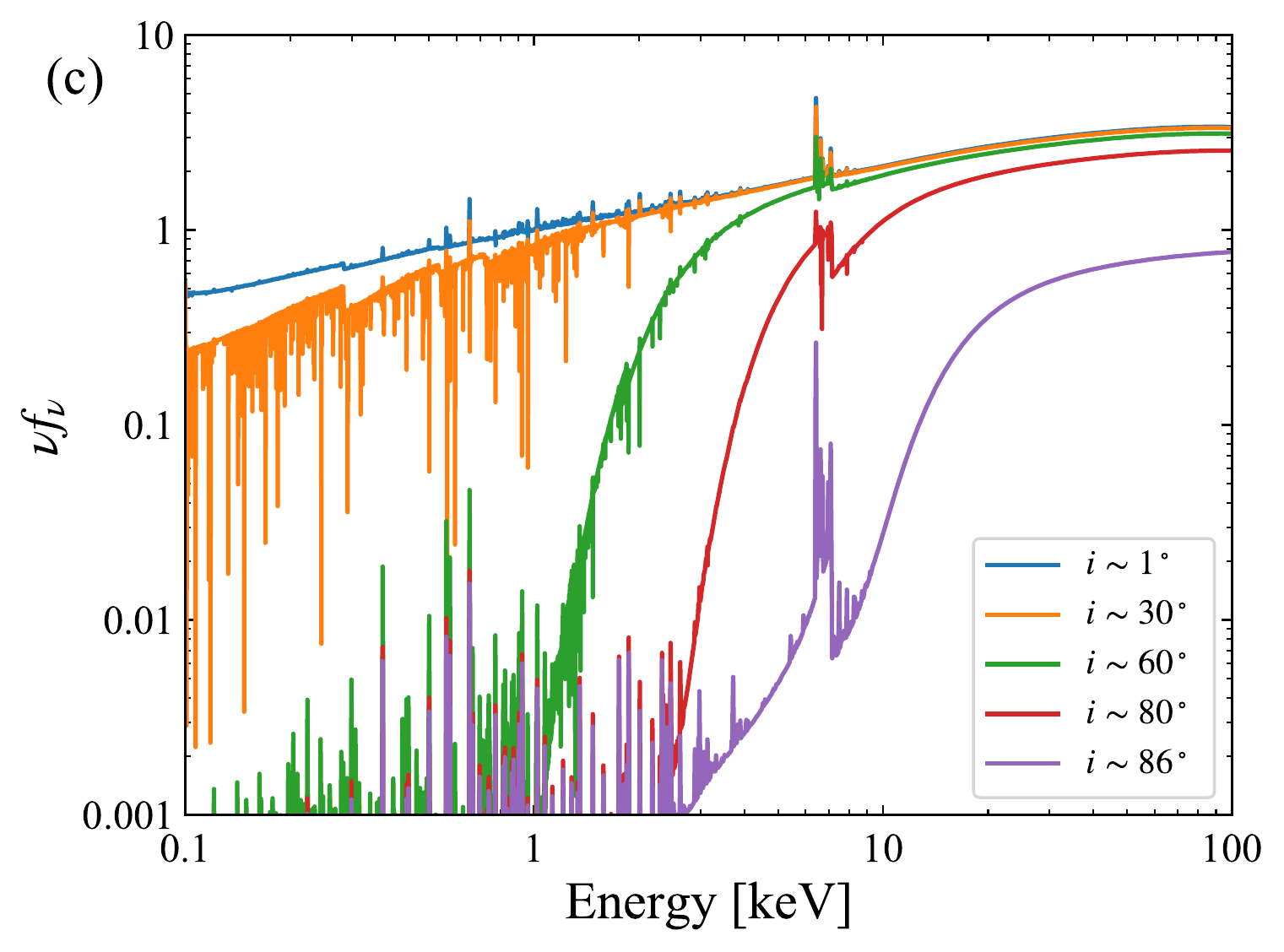}
\caption{
The simulated X-ray spectral models
in units of $\nu f_\nu$ for inclinations of $i=$ 1, 30, 60, 80, and 86~degrees, from top to bottom.
(a): The transmitted spectra (attenuated incident radiation).
(b): The scattered spectra, which contains Compton scattered incident radiation and diffuse emission from all the cells.
(c): The total (transmitted + scattered) spectra.
}
\label{fig_outsed}
\end{figure*}

Figure~\ref{fig_outsed}(a), (b), (c) plot examples of the
transmitted, scattered, and total spectra, respectively, for five
inclination angles ($i=$ 1, 30, 60, 80, and 86~degrees). The spectra are
characterized by many absorption line and edge features and emission
lines. The transmitted spectra are absorbed by materials in the line
of sight, and show many absorption lines by ionized gas in the
0.1--2~keV bandpass at low to medium inclination angles, whereas the
absorption features are very weak at
$i=1$ degree
because of the low
line-of-sight hydrogen column density. 
These results indicate
that, even if the presence of warm absorbers is universal in AGNs,
they can be directly observed only in a certain range of inclination angles.
Ionized emission lines such as
hydrogen and helium-like oxygen lines at $\sim$0.654~keV and 0.56~keV,
respectively, are seen in the scattered spectra in all inclination
angles.
In addition to the emission lines from mildly to highly ionized gas (e.g., \ion{O}{8}~Ly$\alpha$ and \ion{O}{7}~He$\alpha$ lines), 
iron K$\alpha$ fluorescence lines at $\sim$6.4~keV from cold matter is
also produced.

\subsection{\ion{O}{8}~Ly$\alpha$ Intensity Distribution}

\begin{figure*}
\plottwo{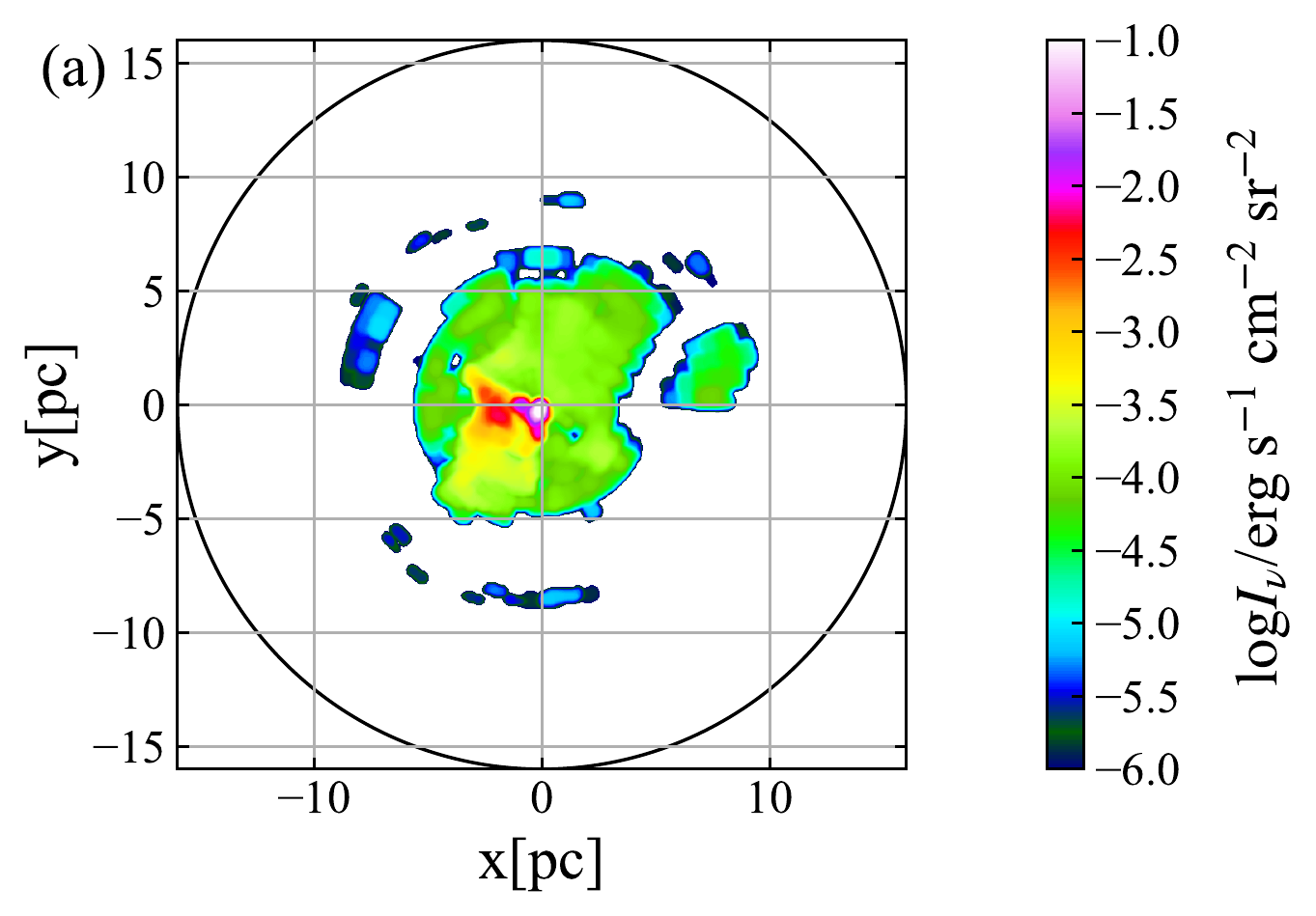}{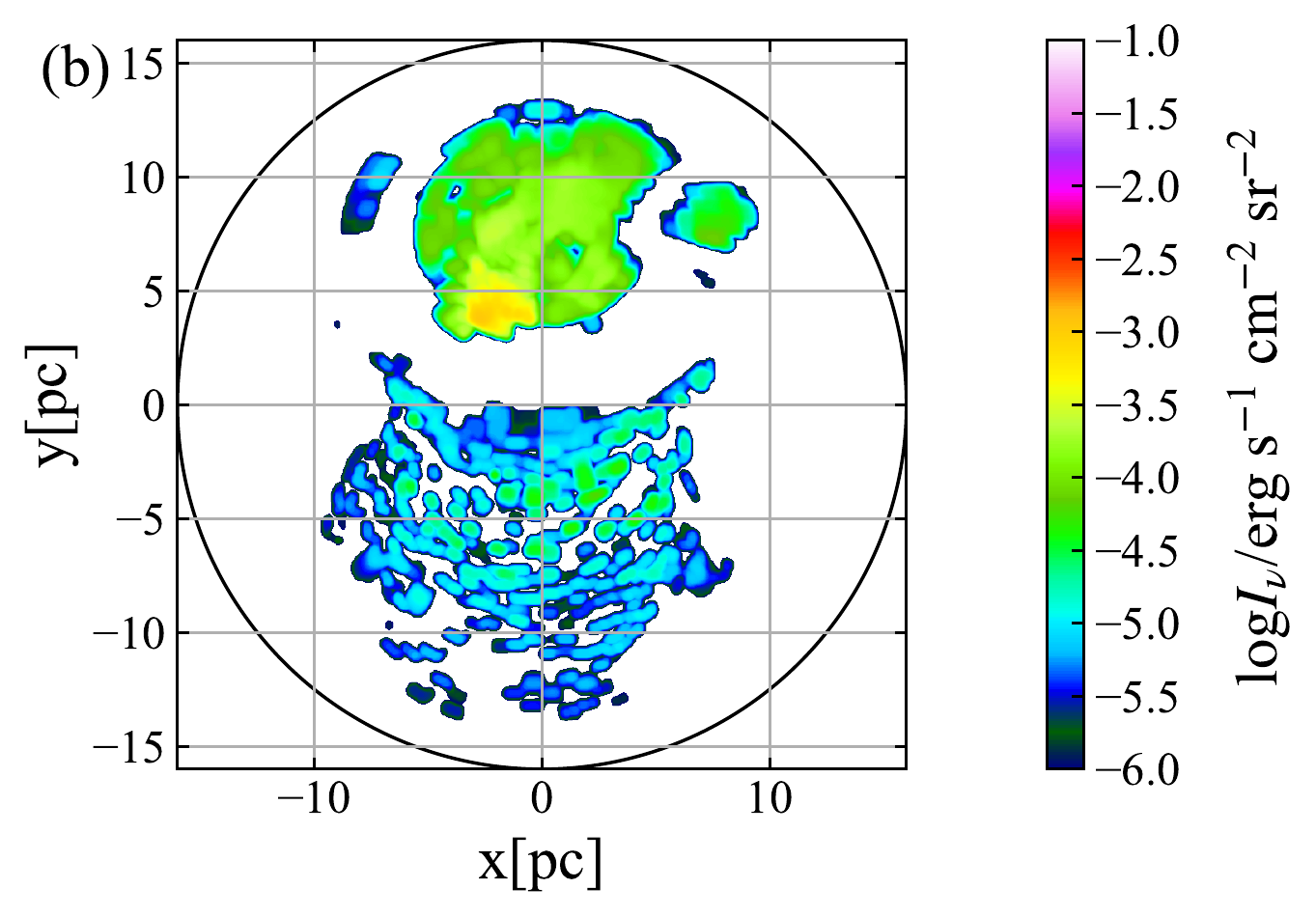}
\caption{
(a): Surface brightness distribution of \ion{O}{8}~Ly$\alpha$ at 0.654~keV for the inclination angle $i = 0^\circ$. 
(b): Same as (a) but for $i = 30^\circ$. 
}
\label{fig-map}
\end{figure*}

The Cloudy simulations enable us to investigate the
spatial distribution of line intensity emitted by the irradiated gas.
To make a surface brightness map
that could be compared with a virtual high-spatial resolution observation,
we integrate the line intensity along the lines of
sight, taking into account absorption in each cell
according to the optical depth.
Figure~\ref{fig-map}(a) and (b) plot the surface brightness maps at 0.654~keV
(corresponding to Ly$\alpha$ of \ion{O}{8}) viewed from $i=0^\circ$
and 30$^\circ$, respectively. 
As noticeable from the map at $i=30^\circ$, the lines
produced in the upper side of the outflow are dominant in the observed spectrum
because
those in the bottom side are largely absorbed by the intervening gas.
The gap at $y =$ 0--5~pc is due to the absorption
by the optically thick, inner torus region.
The \ion{O}{8} distribution shows a conical shape similar to [\ion{O}{3}]
5007~\AA\ \citep{Wada2018b}, although the \ion{O}{8}~Ly$\alpha$
emitting region is located closer to the SMBH than the [\ion{O}{3}]
5007~\AA. 

\section{Comparison with Observations}
\label{sec4}

\begin{figure}
\plotone{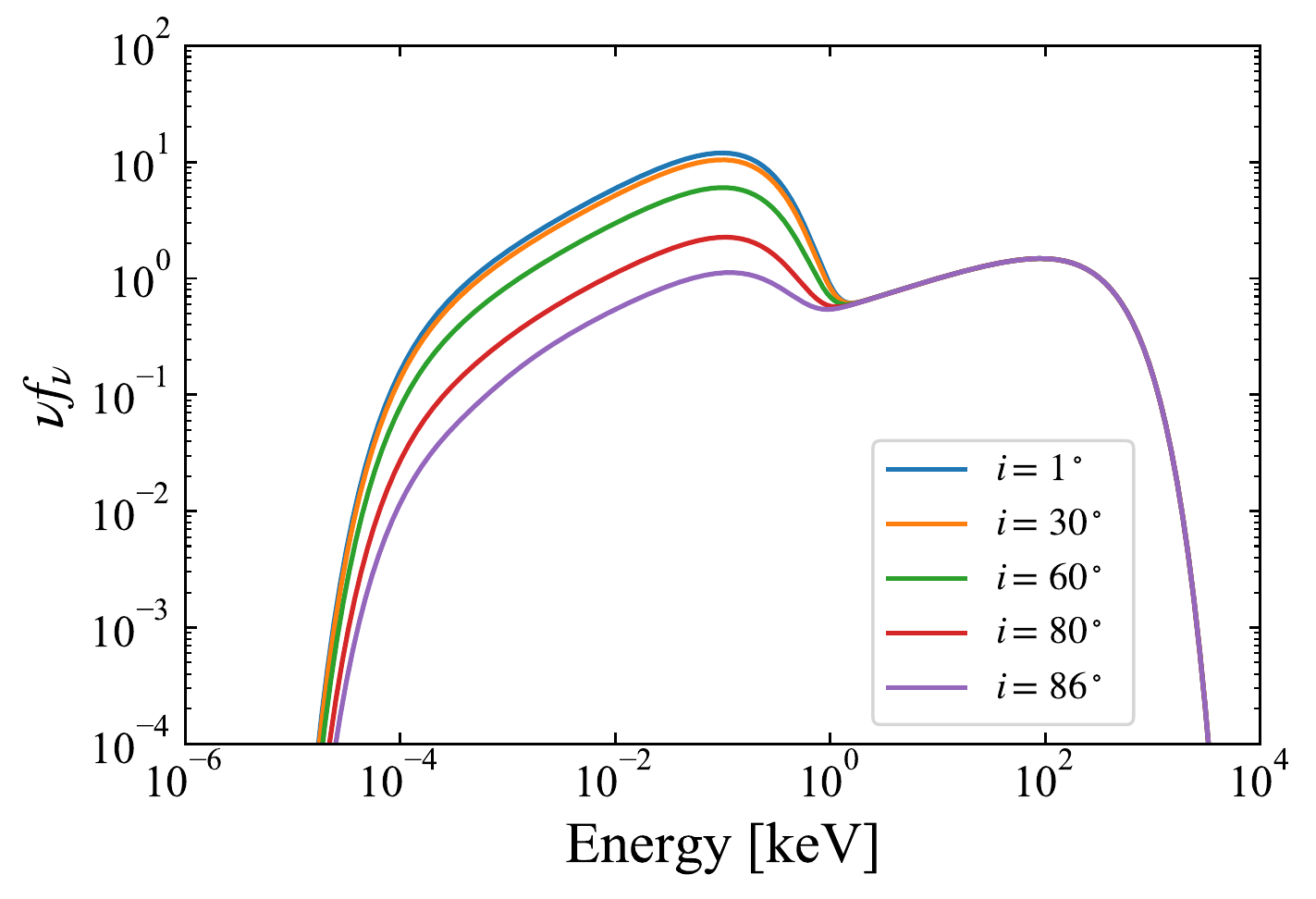}
\caption{
  Incident SED models of NGC~4051 at the innermost grids for
  inclination angle of $i=$ 1, 30, 60, 80, and 86~degrees (from top to
  bottom).
The vertical axis has an arbitrary unit of $\nu f_\nu$.}
\label{fig-insed2}
\end{figure}

\begin{figure*}
\epsscale{1.0}
\plottwo{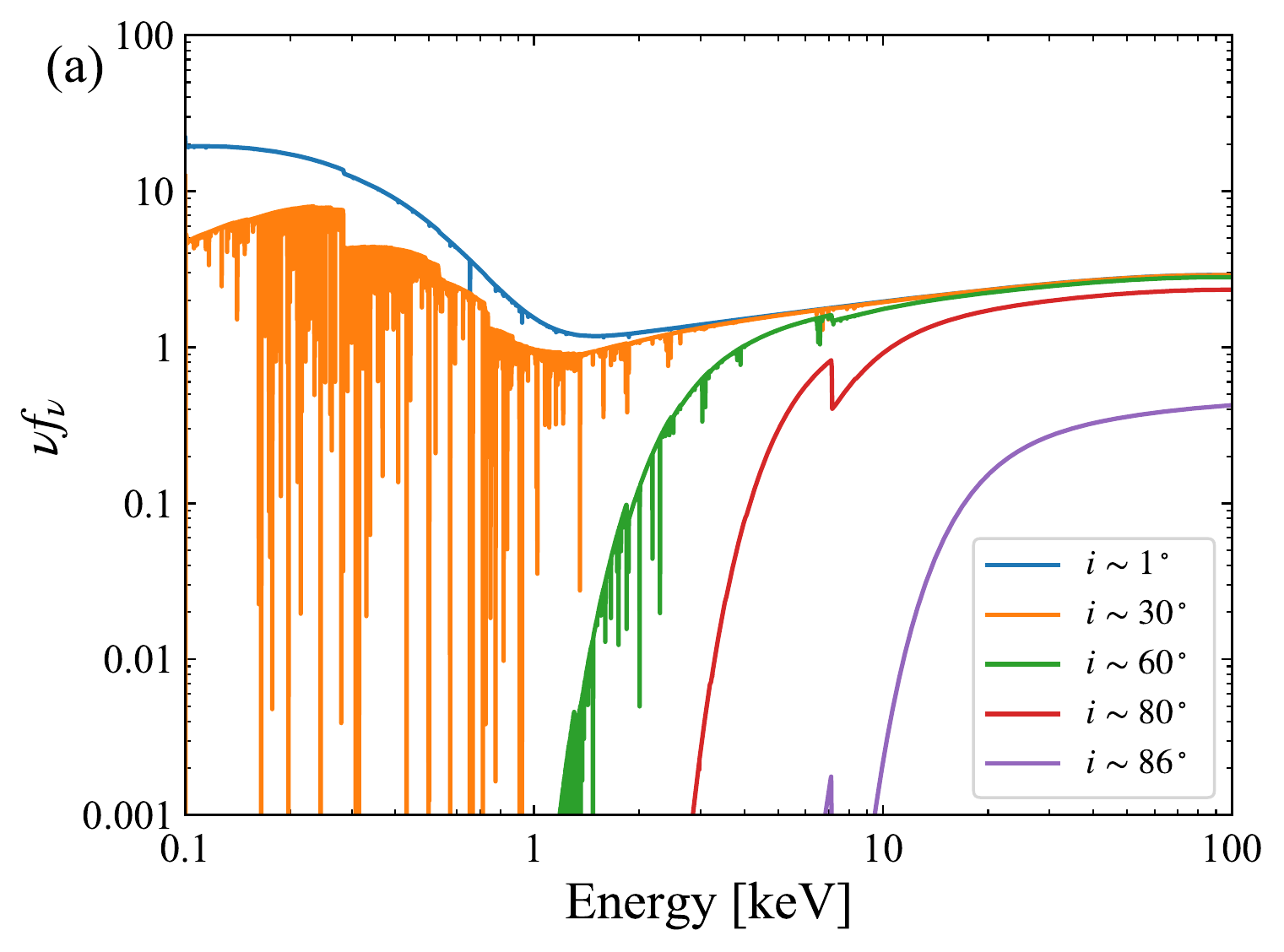}{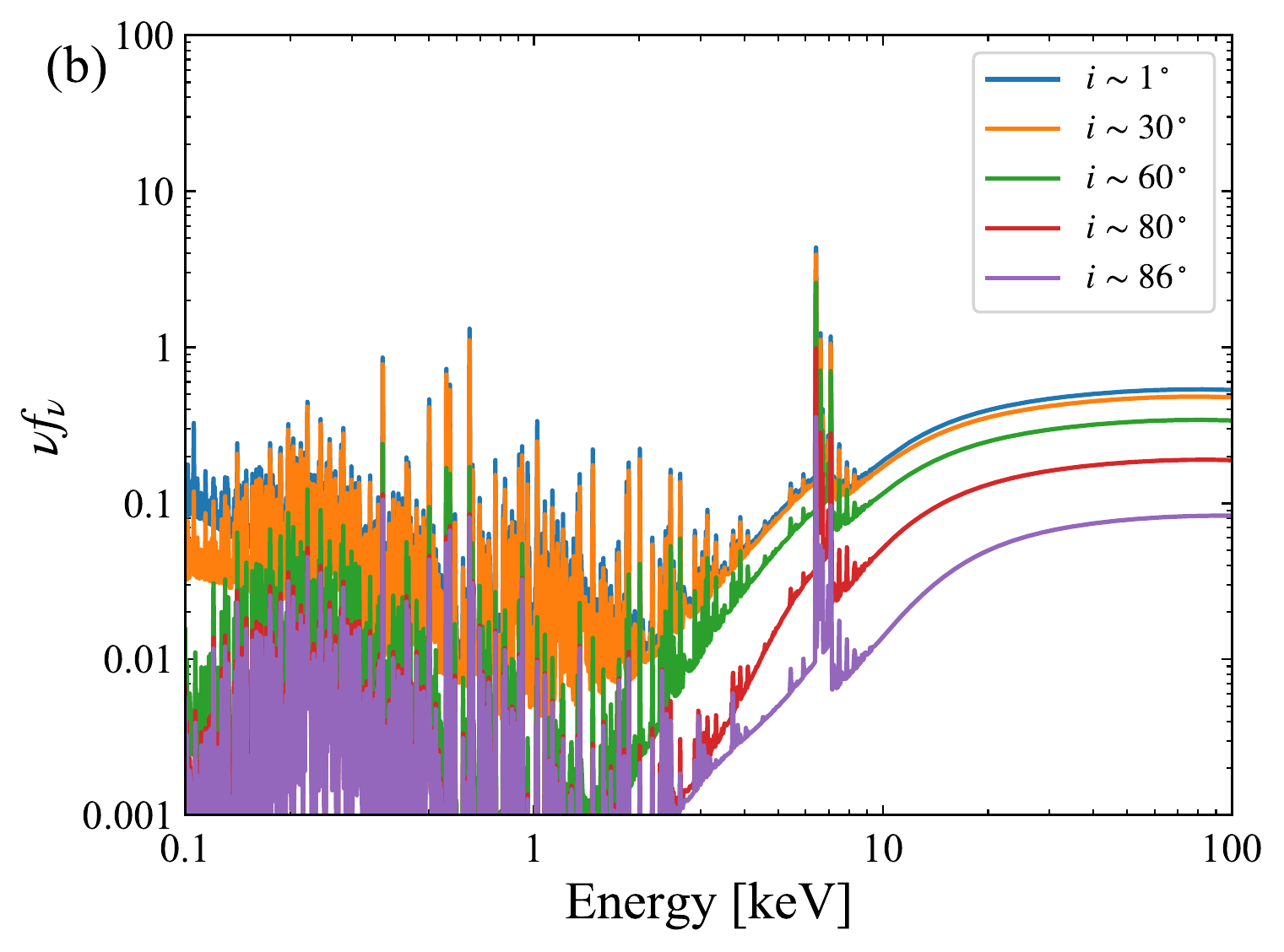}
\epsscale{0.5}
\plotone{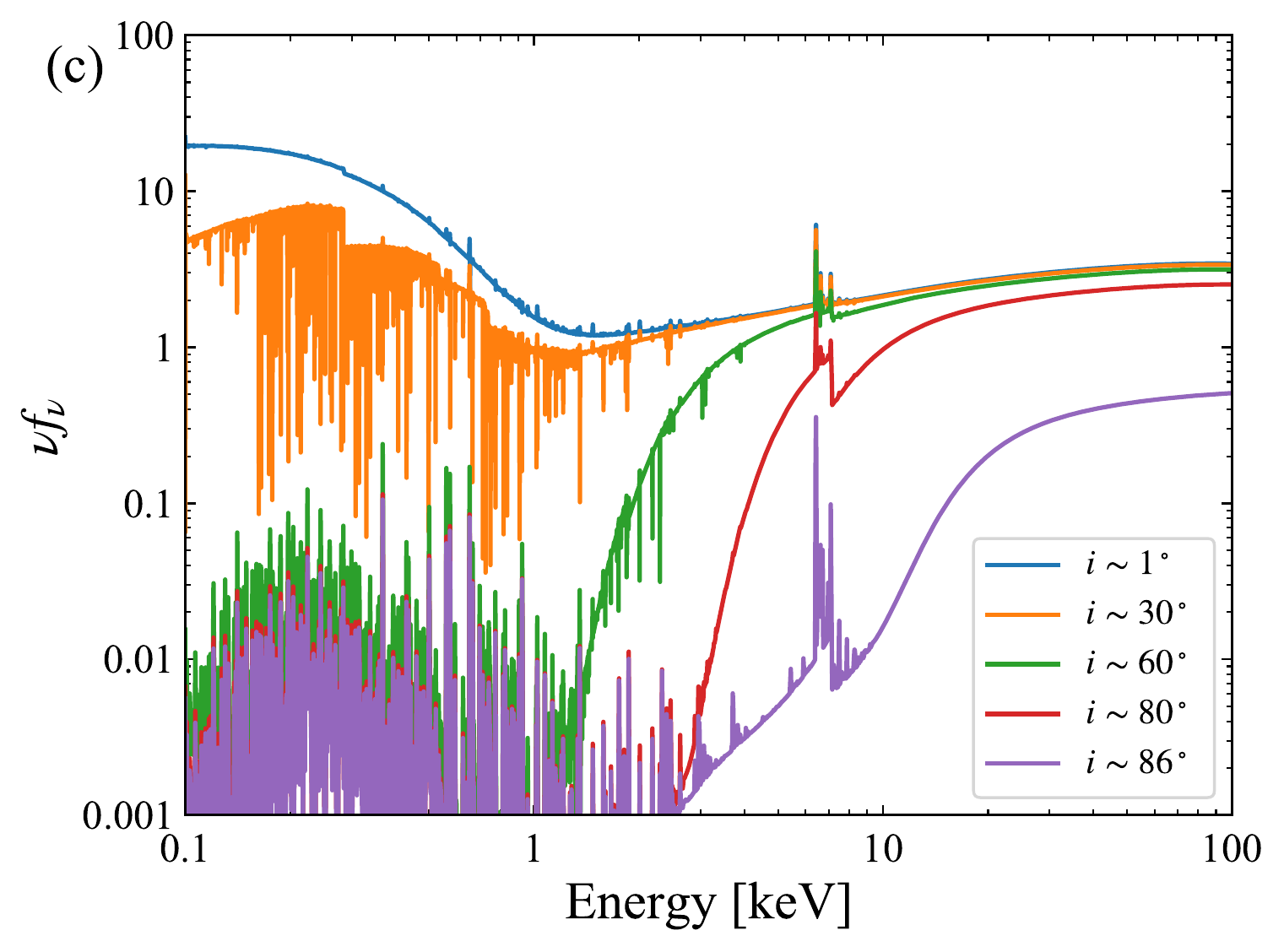}
\caption{
The simulated X-ray spectral models for NGC~4051
in units of $\nu f_\nu$ for inclinations of $i=$ 1, 30, 60, 80, and 86~degrees, from top to bottom.
(a): The transmitted spectra (attenuated incident radiation).
(b): The scattered spectra, which contains Compton scattered incident radiation and diffuse emission from all the cells.
(c): The total (transmitted + scattered) spectra.
}
\label{fig_outsed2}
\end{figure*}

In this section, we compare the results of our mock observations of a
radiation-driven fountain with actual observations, regarding the
properties of the ionized absorbers and X-ray spectra. For the
comparison object with a low inclination, where the warm absorbers are
directly observed, we have selected NGC~4051.
For a high inclination case, we show the comparison
with the Circinus galaxy (Seyfert 2) in Appendix~\ref{A1.2}.
\citet{Wada2016} adopted the same AGN parameters as those of the Circinus galaxy, 
($\log M_\mathrm{BH}/M_\odot$, $\log L_\mathrm{bol}$, $\lambda_\mathrm{Edd}$)
= ($6.3$, $43.7$, $0.2$).

In order to make a precise comparison with NGC~4051,
we performed the \textsf{Cloudy} 
simulations by changing the input SED to
that more suitable for NGC~4051. This is mainly because NGC~4051 
shows a strong soft excess component
in the 0.4--1~keV band over the power-law component
\citep[e.g.,][]{Nucita2010,Mizumoto2017,Ogawa2021}.
To find an appropriate SED model of NGC~4051,
we performed a 
SED fitting with the equation~\ref{eqn:agncon} using the time-averaged data of
XMM-Newton \citep{Jansen2001} EPIC/pn and optical monitor (OM). 
We derived $T_\mathrm{BB} = 2\times10^6$~K, $\alpha_\mathrm{X} = -0.7$, and $\alpha_\mathrm{OX} = -1.3$.
The other parameters are fixed at the values adopted in the previous simulations.
Since 
$T_\mathrm{BB}$ is 20 times larger than that in \citet{Wada2018b},
the prominent soft excess component appears in the 0.4--1~keV band.
The ionizing luminosity $L_\mathrm{ion}$ (0.0136--13.6 keV) and  
the ionization parameters $\xi$ are increased by a factor of $\sim$5 (i.e., $\log \xi$ is increased by $\sim$0.7) compared with the original SED.
We assumed that the contamination from the host galaxy
has negligible effects on the OM data.
Figures~\ref{fig-insed2}
shows input SED models
at five
inclination angles ($i=$ 1, 30, 60, 80, and 86~degrees).
The output spectra of the transmitted, scattered, and total components are
plotted in Figure~\ref{fig_outsed2}.

Note that the SED assumed in the \textsf{Cloudy} simulation for NGC~4051 are not strictly the same as those assumed for the Circinus galaxy \citep{Wada2016} (see Figures~\ref{fig-insed} and \ref{fig-insed2}). 
However, as \citet{Wada2015} showed, the structure of the radiation-driven fountain will not change significantly as long as the Eddington ratio is similar 
(in the range of 0.2--0.3). Therefore, we here try to compare the fountain model and the X-ray warm absorbers in NGC~4051.

\subsection{Ionized Absorbers}
\label{4-1}

\begin{figure}
\plotone{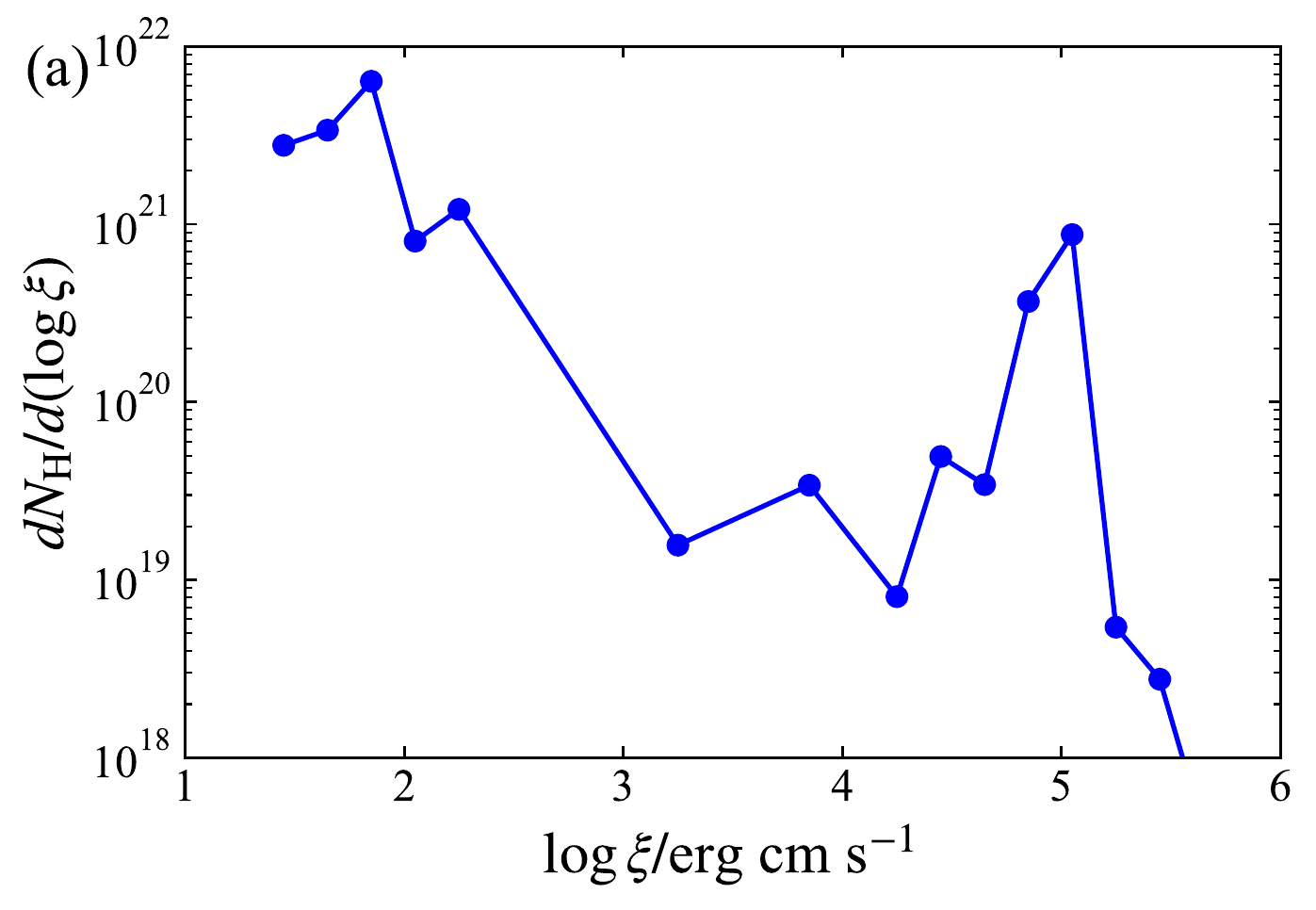}
\plotone{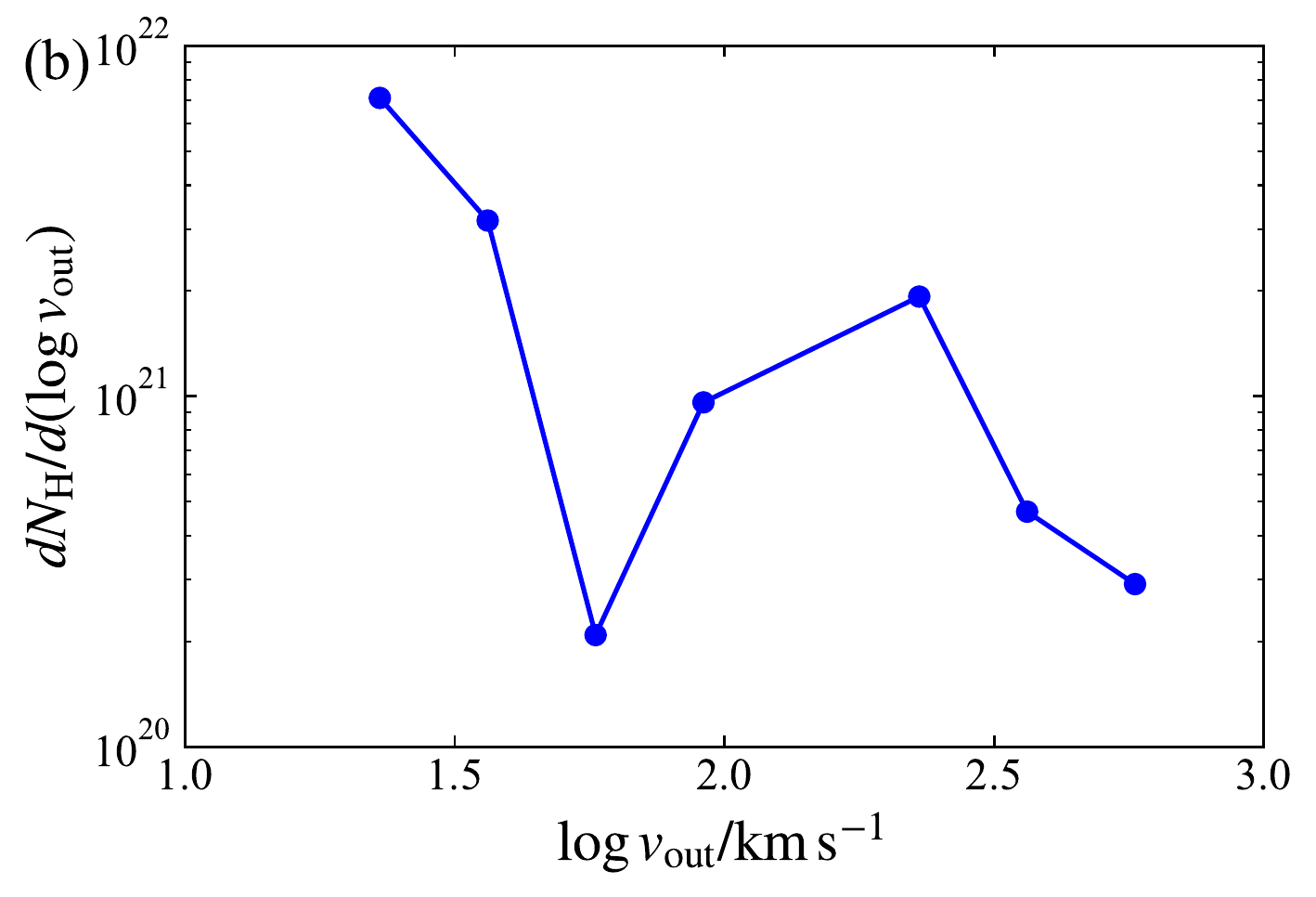}
\caption{
Model predicted absorption measure
distributions (AMDs) for $i=$~30~degrees, which corresponds to the best-fit inclination angle of NGC~4051. 
(a): that of $\log \xi$.
(b): that of $\log v_\mathrm{out}$.
}
\label{fig-amd}
\end{figure}

We investigate the absorption measure distribution (AMD)
of $\xi$ and $v_\mathrm{out}$, which
provides quantitative information on the ionization and velocity
structures of an outflow.
Several attempts have been made to derive the AMDs
utilizing high-resolution X-ray spectra of type-1 AGNs.
In most cases, the AMDs have been found to be
multi-peaked, indicating the presence of multi-phase ionized
absorbers along the line of sight
\citep[e.g.,][]{Holczer2007,Adhikari2019}.
We adopt the definition by \cite{Holczer2007} for the AMD
as a function of the parameter $x$: 
\begin{eqnarray}
    \mathrm{AMD}\left(x\right) \equiv \frac{dN_\mathrm{H}}{d\left(\log x\right)}.
\end{eqnarray}
Figure~\ref{fig-amd} displays the AMDs of ionization parameter $\xi$
and of outflow velocity $v_\mathrm{out}$ obtained at $i=30^\circ$ from
our simulation data.
As noticed, they spread over wide ranges of the parameters.
In NGC~4051, ionization parameters and outflow velocities 
of multiple warm absorbers are consistent with our AMDs
($\log \xi = 1.4-5$ and $\log v_\mathrm{out} = 2.3-2.8$ have been actually observed:
\citealt{Krongold2007,Steenbrugge2009,Lobban2011,Pounds2011,King2012,Silva2016,Mizumoto2017}).
These slow ionized absorbers whose $v_\mathrm{out}$ are a few hundreds km~s$^{-1}$ may be interpreted as the origin 
of the warm-absorber outflow from the torus scale \citep{Blustin2005}.

Here we make detailed comparison with the results of
\citet{Lobban2011}, who reported that four distinct ionization zones
  with outflow velocities of $<$1000~km~s$^{-1}$
were required to reproduce the Chandra/HETG data of NGC~4051 observed in 2008 November.
Our AMD gives a reasonably good description of
their zones 3a, 3b, and 4, whose warm absorber parameters are 
($\log N_\mathrm{H}$, $\log \xi$, $\log v_\mathrm{out}$) = 
($21.0$, $2.16$,  $2.74$), ($20.7$, $1.96$, $2.91$), and ($21.4$, $2.97$, $2.85$), respectively.
On the other hand, zones 1 and 2,
which have ($\log N_\mathrm{H}$, $\log \xi$, $\log v_\mathrm{out}$) = 
($20.5$, $-0.86$, $2.25$) and ($20.2$, $0.60$, $2.34$), respectively,
are not reproduced in our model; 
their ionization parameters are much lower than those
in our model. To reproduce these low ionization, low column-density
material, numerical simulations with higher spatial resolution may be needed.
Assuming $L_\mathrm{ion} = 10^{43}$~erg~s$^{-1}$, $\xi =
L_\mathrm{ion}/n_\mathrm{H}r^2 = 1$, and $r = $10~pc, where $\log
v_\mathrm{out}$ is $\sim$2 (Figure~\ref{fig_reformed}(c)),
$n_\mathrm{H} \sim 10^4$~cm$^{-3}$ and therefore $N_\mathrm{H} = n_\mathrm{H}
\Delta r \sim 10^{22}$~cm$^{-2}$ for $\Delta r$ = 0.25~pc, 
which is the
resolution along the radial direction in this work\footnote{The
spatial resolution of the original hydrodynamic grid data is
0.125~pc \citep{Wada2016}}. To reproduce the observed low column densities
of zones 1 and 2, clumpy gas whose size is 
1.5--2 orders of magnitude smaller than the current resolution 
must be considered.

In addition to these relatively slow components, 
faster ones with $v_\mathrm{out} > 4000$~km~s$^{-1}$ are also detected
\citep{Steenbrugge2009,Lobban2011,Pounds2011,Silva2016,Mizumoto2017}. The latter cannot be reproduced by
our current model.
This is because the model does not include
any matter
inside the innermost numerical grid (i.e., $r =0.125$~pc), where the
  escape velocities for a black hole mass of $2 \times 10^6 M_\odot$ are
  $>$370~km~s$^{-1}$. Assuming that the outflow velocity corresponds
to the escape velocity, these faster outflows should be launched at 
regions closer to the SMBH than the torus, 
such as the line-driven wind \citep[e.g.,][for UFOs]{Nomura2017,Nomura2020}.
In fact, recent radiation-hydrodynamic simulations suggest that, 
if the dusty gas of a thin disk continues to the inner 0.01~pc under the non-spherical UV radiation field\footnote{The dust sublimation radius $r_\mathrm{sub}$ assuming spherical radiation 
is given by the formula of \citet{Barvainis1987},
$r_\mathrm{sub} = 1.3 \left( L_\mathrm{UV}/10^{46} \ \mathrm{erg \ s}^{-1} \right)^{0.5} \left( T_\mathrm{sub}/1500 \ \mathrm{K} \right)^{-2.8}$~pc, 
where $L_\mathrm{UV}$ is the UV luminosity
and $T_\mathrm{sub}$ is the dust sublimation temperature.
Adopting $L_\mathrm{UV} = 2\times10^{43}$~erg~s$^{-1}$ and
$T_\mathrm{sub} = 1500$~K, we obtain $r_\mathrm{sub}$=0.05~pc for
NGC~4051.
However,
if anisotropic radiation from the accretion disk
is taken into account, $r_\mathrm{sub}$ can be smaller than this value at
the surface of a thin disk\citep[see e.g.,][]{Kawaguchi2010}.},
the gas can be 
blown away with outflow velocities of $>$1000~km~s$^{-1}$ (Kudoh, Y., et al., in preparation). 
This fast outflow may be able to explain the properties of the observed fast warm absorbers.
Another possibility is the magnetically driven outflows \citep[e.g.,][]{Fukumura2010,Fukumura2014}, but
simulations of magnetic winds depend on the (currently unknown) magnetic field configuration.

\subsection{X-ray Spectrum}
\label{sec4.2}

\begin{figure}
\plotone{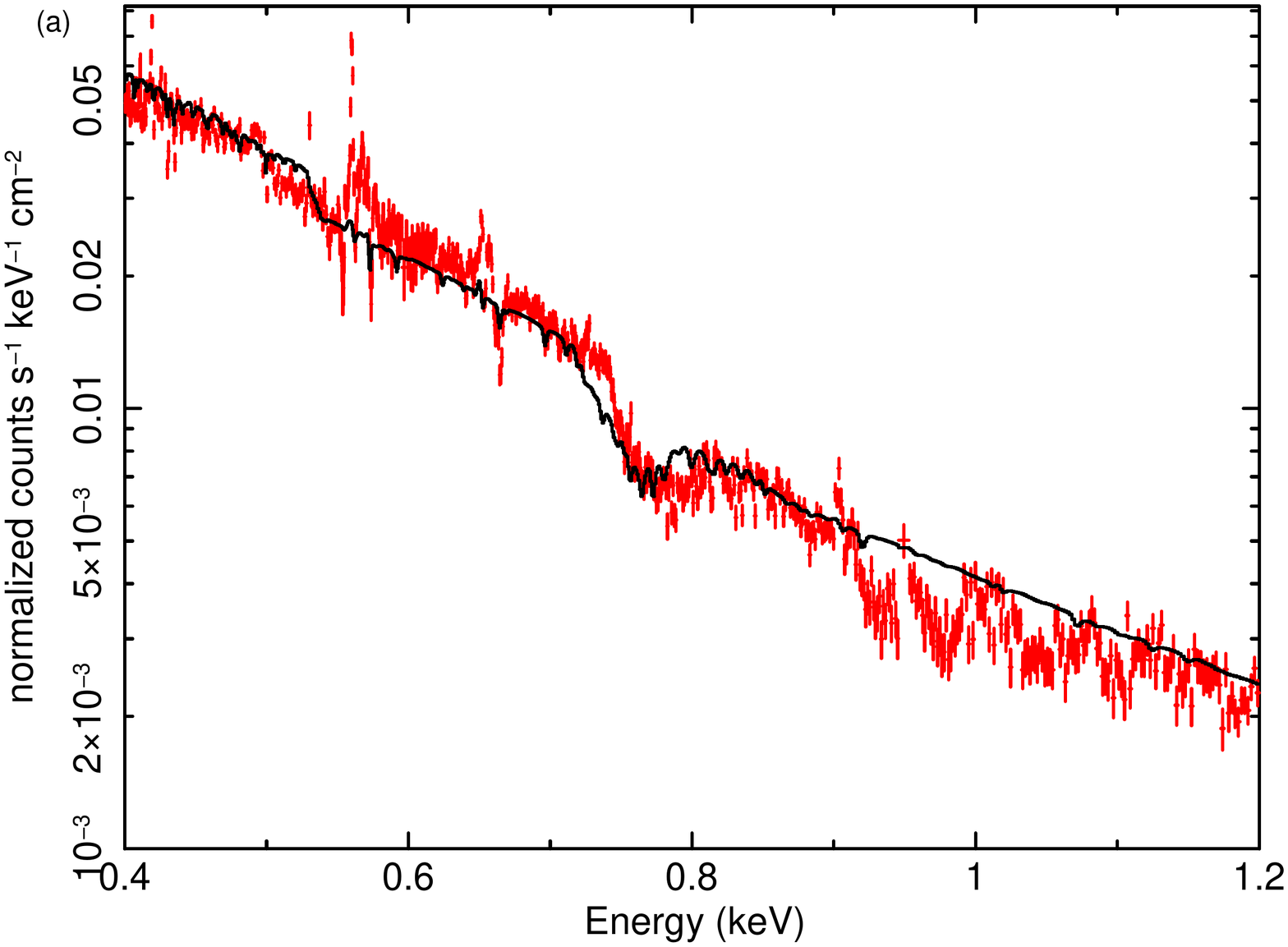}
\plotone{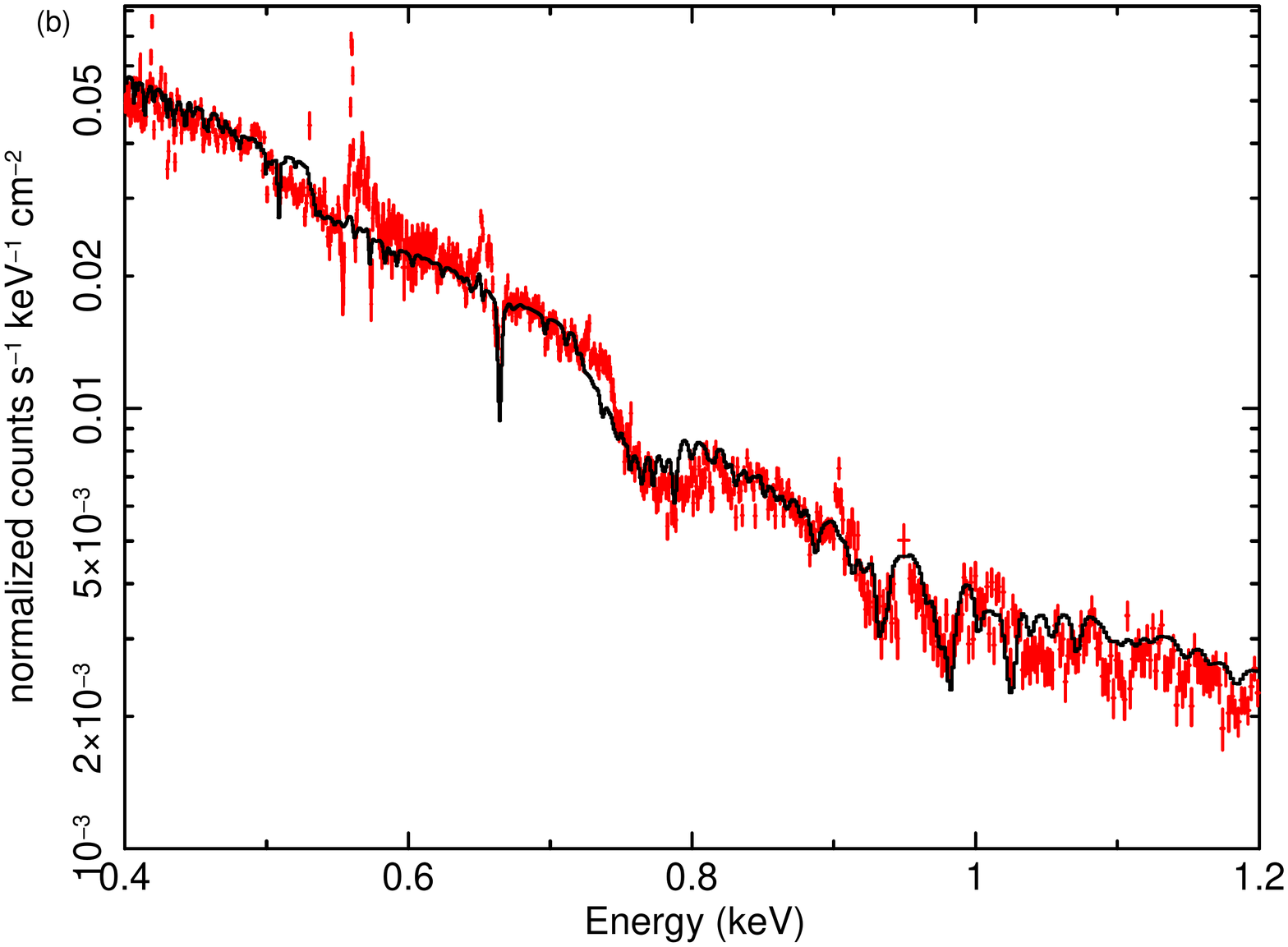}
\caption{
(a): The spectrum of NGC~4051
observed with \textit{XMM-Newton}/RGS (red crosses). It is 
folded with the energy responses but is corrected for effective area.
The solid curve represents the best-fit model based on our \textsf{Cloudy}
simulations. 
(b): Same as (a), but fit with the model including
a fast warm-absorber component.}
\label{fig-spec}
\end{figure}

We compare our simulated spectra with the high energy-resolution spectrum
of NGC~4051 observed with XMM-Newton/RGS\citep{den_Herder2001}. 
Our purpose is to check to what extent our simulations
may produce the observed spectrum, in order to obtain insights on the
AGN structure. Efforts to find spectral models that are fully
consistent with the observed spectrum are beyond the scope of this
paper.

We analyzed the archival data observed in 2009 May--June (observation ID:
0606320101, 0606320201, 0606320301, 0606320401, 0606321301,
0606321401, 0606321501, 0606321601, 0606321701, 0606321801,
0606321901, 0606322001, 0606322101, 0606322201, 0606322301) using the
Science Analysis Software (SAS) version~17.0.0 and calibration files
(CCF) released on 2018 June 22. 
These observation epochs are different from that of the Chandra/HETG
observation discussed in the previous subsection \citep{Lobban2011}.
The RGS data were reprocessed with
\textsc{rgsproc}. We stacked all the data to create the highest
signal-to-noise time-averaged spectrum.

We fit the spectrum in the 0.4--1.6~keV range with a model based on our
simulations using the Cash statistic \citep{Cash1979}.
In the Xspec terminology, the model is represented as 
``\textsf{phabs} $\times$ (\textsf{mtable\{fountain\_T.fits\}} $\times$
\textsf{zcutoffpl1 + zcutoffpl2} $+$ \textsf{atable\{fountain\_S.fits\}})''. 
The \textsf{phabs} term represents the Galactic absorption fixed at
$N_\mathrm{H} = 1.20 \times 10^{20}$~cm$^{-2}$ (the total Galactic \ion{H}{1}
and H$_2$ value given by \citealt{Willingale2013}). 
We use the \textsf{zcutoffpl1} and \textsf{zcutoffpl2} to represent the first and second terms in Equation~\ref{eqn:agncon}, respectively.
The table model
\textsf{atable\{fountain\_S.fits\}} represents scattered spectra of our simulation, whereas \textsf{mtable\{fountain\_T.fits\}}
takes into account the absorption to the transmitted component.
The free parameters are the cutoff energy and normalization of 
\textsf{zcutoffpl1}, the normalization of \textsf{zcutoffpl2},
 and the inclination and azimuthal angles in the two table models.
The normalization of \textsf{atable\{fountain\_S.fits\}} is linked to that of \textsf{zcutoffpl2},
the photon index of \textsf{zcutoffpl1} is fixed at 1.5,
and the photon index and cutoff energy of \textsf{zcutoffpl2} are fixed at 1.7 and 300~keV, respectively.

Figure~\ref{fig-spec}(a) plots the observed spectrum of NGC~4051
and the best-fit model
folded with the energy responses
(corrected for the effective area).
We obtain the best-fit inclination angle of $\approx$30$^\circ$ ($C/$dof $= 9621/2304$).
It is seen that our model 
reproduces many weak absorption features, particularly at energies
below 0.6~keV. As mentioned above, however, it cannot reproduce the
strong absorption features at $\geq$0.9~keV, which are produced by
fast warm absorbers with $v_\mathrm{out} \sim$ 4000--6000~km~s$^{-1}$ \citep{Pounds2011,Silva2016,Mizumoto2017}.
If a fast component with $\log \xi \sim 2.6$, $\log N_\mathrm{H} \sim 22.2$, $v_\mathrm{out} \sim 4700$~km~s$^{-1}$ is added to our model,
these absorption features can be reproduced (Figure~\ref{fig-spec}(b)) and the fit is much improved ($C/$dof $= 7944/2301$).

It is also noteworthy that the simulated spectra have weaker emission lines
than the observed data; 
the equivalent width of \ion{O}{8}~Ly$\alpha$
  is by a factor of $\sim$8
smaller than that of observed result \citep{Mizumoto2017}.
These emission lines may also come from
regions inside the space where the hydrodynamic simulations were
performed (i.e., $<$0.125~pc). In fact, 
the line width of \ion{O}{8}~Ly$\alpha$,
$\sigma = 0.1$~\AA\ \citep{Mizumoto2017} (or 1600~km~s~$^{-1}$),
is too broad to be produced by our model.
Fast warm absorbers launched at $r<$0.125~pc 
(see section~\ref{4-1}) might produce these emission lines. 

Our model also shows some discrepancies with the observed spectrum
in the Fe M-shell unresolved transition array (UTA) feature 
around 0.7~keV. This is mainly because the
the column density and outflow velocity of 
the best-fit model do not correctly represent those
It is also noticeable that the model 
overestimates the flux around 0.5--0.53 keV.
We infer that this is because the model does not well reproduce the K-edge
feature 
from H-like carbon ions at 0.49~keV\citep{NIST_ASD}.
At the best-fit inclination angle, which is mainly determined by the UTA feature, carbon atoms are almost fully ionized and hence the edge structure is very weak.
Here, one should also note that the 
three dimensional substructures of the absorbers could be highly time variable.
In fact, 
as \citet{Schartmann2014} showed (see their Figure~6), the observed SEDs, especially in the short wavelengths, varies more than two orders of magnitude
due to the absorption for a given inclination angle over 0.1--1~Myr.

\section{Conclusions}
\label{sec5}

\begin{enumerate}

\item We have investigated the ionization state and X-ray spectra of
  the radiation-driven fountain model around a low-mass AGN
  \citep{Wada2016}, utilizing the \textsf{Cloudy} code.

\item The model makes multi-phase ionization structure around the
  SMBH, as is actually observed in ionized outflows of AGNs. The
  simulated spectra show many absorption features and emission lines
  of ionized materials, which highly depends on inclination angle.

\item 
  We compare our mock observations with the actual spectrum of NGC~4051. 
  Although the radiation-driven fountain model accounts for warm
  absorbers with low ionization parameters ($\log \xi \lesssim 2$) and
  low outflow velocities of $>$a few hundred km~s$^{-1}$, it cannot
  reproduce faster ones with a few thousands km~s$^{-1}$. Also, our
  simulated spectra show a weaker and narrower
  \ion{O}{8}~Ly$\alpha$ emission line
  than the observed one by factors of $\sim$8 and
  $\sim$2, respectively.

\item 
  These results suggest that additional components originated 
  from the region closer to the SMBH than the torus via e.g.,
  line-driven winds produced in the accretion disk, must be an important
  ingredient of the warm absorbers in AGNs.

\end{enumerate}

\acknowledgements 

This work has been financially supported by the Grant-in-Aid for JSPS
Research Fellowships 21J13894 (S.O.), for Scientific Research
20H01946 (Y.U.) and 21H04496 (K.W. and Y.U.), 
and for Early-Career Scientists 21K13958 (M.M.).
We are grateful to R. Uematsu for kindly providing
us with the spectra of the Circinus galaxy, which he had analyzed.
M.M. acknowledges the Hakubi project at Kyoto University.
This research has made use of \textsf{Cloudy}.  This research has also made use
of data obtained with \textit{XMM-Newton}, an ESA science mission with
instruments and contributions directly funded by ESA Member States and
NASA, and use of data and software provided by the High Energy
Astrophysics Science Archive Research Center (HEASARC) and the Chandra
X-ray Center (CXC).
The radiation-hydrodynamic simulations were performed on a Cray XC50 supercomputer at the Center for Computational Astrophysics, National Astronomical Observatory of Japan.

\vspace{5mm}
\facilities{XMM-Newton, Chandra.}

\software{HEAsoft \citep[v6.26.1;][]{HEAsoft2014}, SAS \citep[v17.0.0;][]{Gabriel2004}, XSPEC \citep[v12.10.1f;][]{Arnaud1996}, \textsf{Cloudy} \citep[v17.02;][]{Ferland2017}.}

\newpage
\appendix
\restartappendixnumbering

\section{Cold Reflectors in the Radiation-Driven Foutain Model}
\label{A1}

For reference, we also investigate the nature of cold reflectors in
the radiation-driven fountain model, with a particular focus on the
fluorescence Fe~K$\alpha$ line.

\subsection{Fe K$\alpha$ Intensity Distribution}

\begin{figure*}
\plottwo{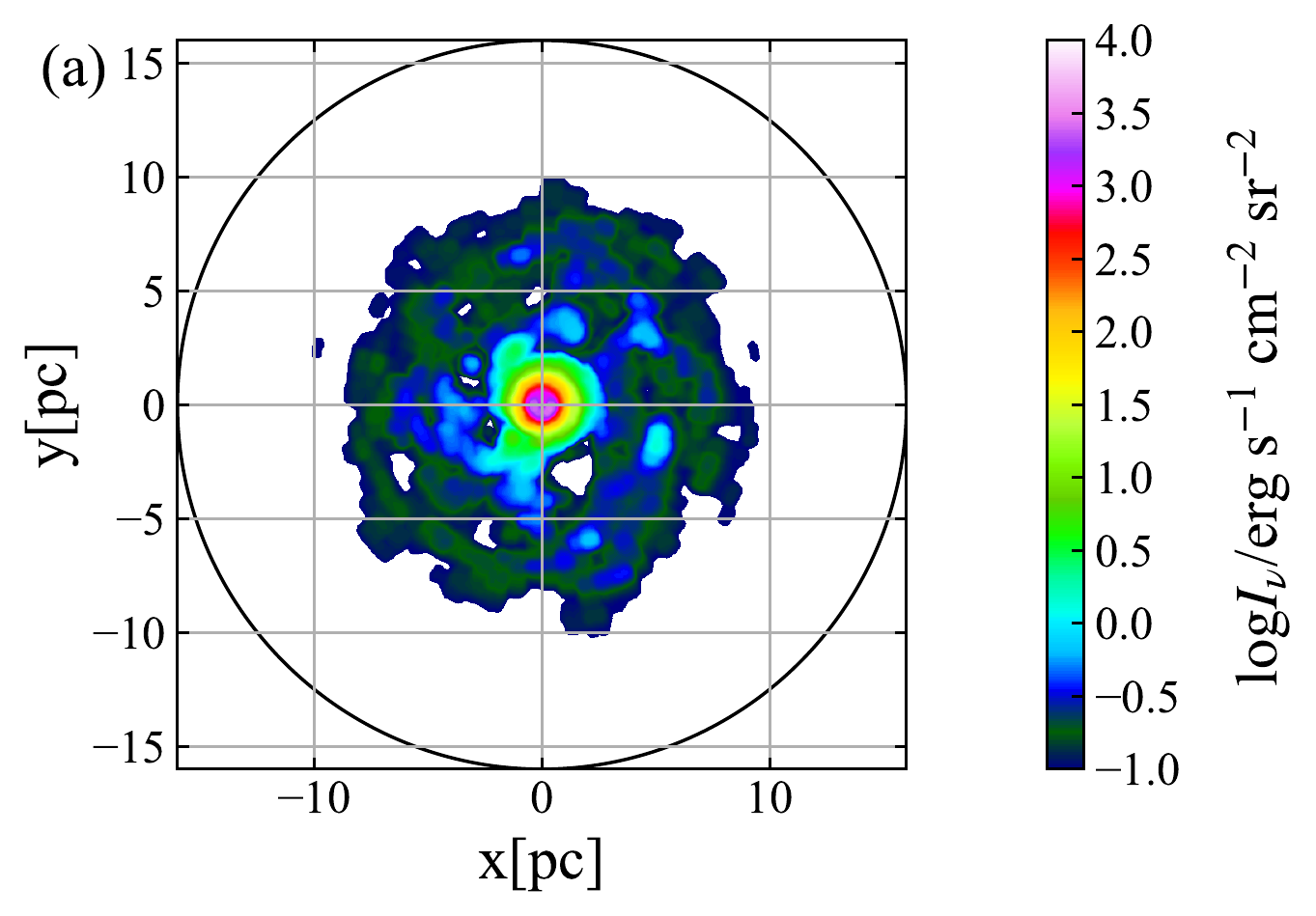}{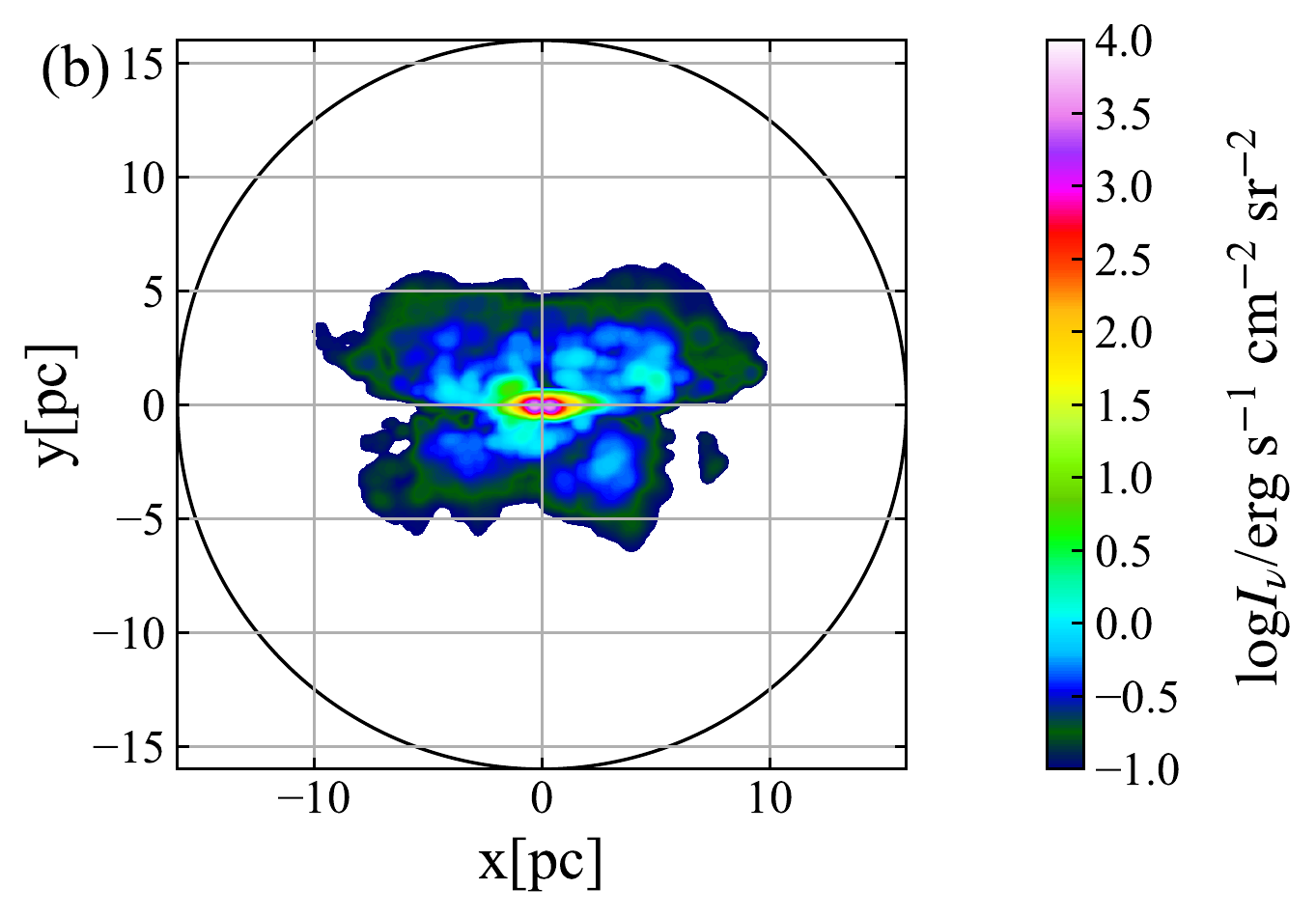}
\caption{
(a): Surface brightness distribution of Fe~K$\alpha$ at 6.4~keV for $i = 0^\circ$. 
(b): Same as (a) but for $i = 80^\circ$. 
}
\label{figA-map}
\end{figure*}

Figure~\ref{figA-map}(a) and (b) plot the surface brightness maps
of lowly ionized Fe~K$\alpha$ lines at 6.4~keV.
When viewed from nearly edge-on,
the Fe~K$\alpha$ lines mainly come from the far side of the inner
torus. This is consistent with the results of a clumpy torus model
\citep[XCLUMPY;][]{Tanimoto2019} studied by \citet{Uematsu2021}, who
investigated the locations of Fe~K$\alpha$ emitting regions with
Monte-Carlo ray-tracing simulations.

\subsection{X-ray Spectrum: Comparison with Circinus Galaxy}
\label{A1.2}

\begin{figure}
\plotone{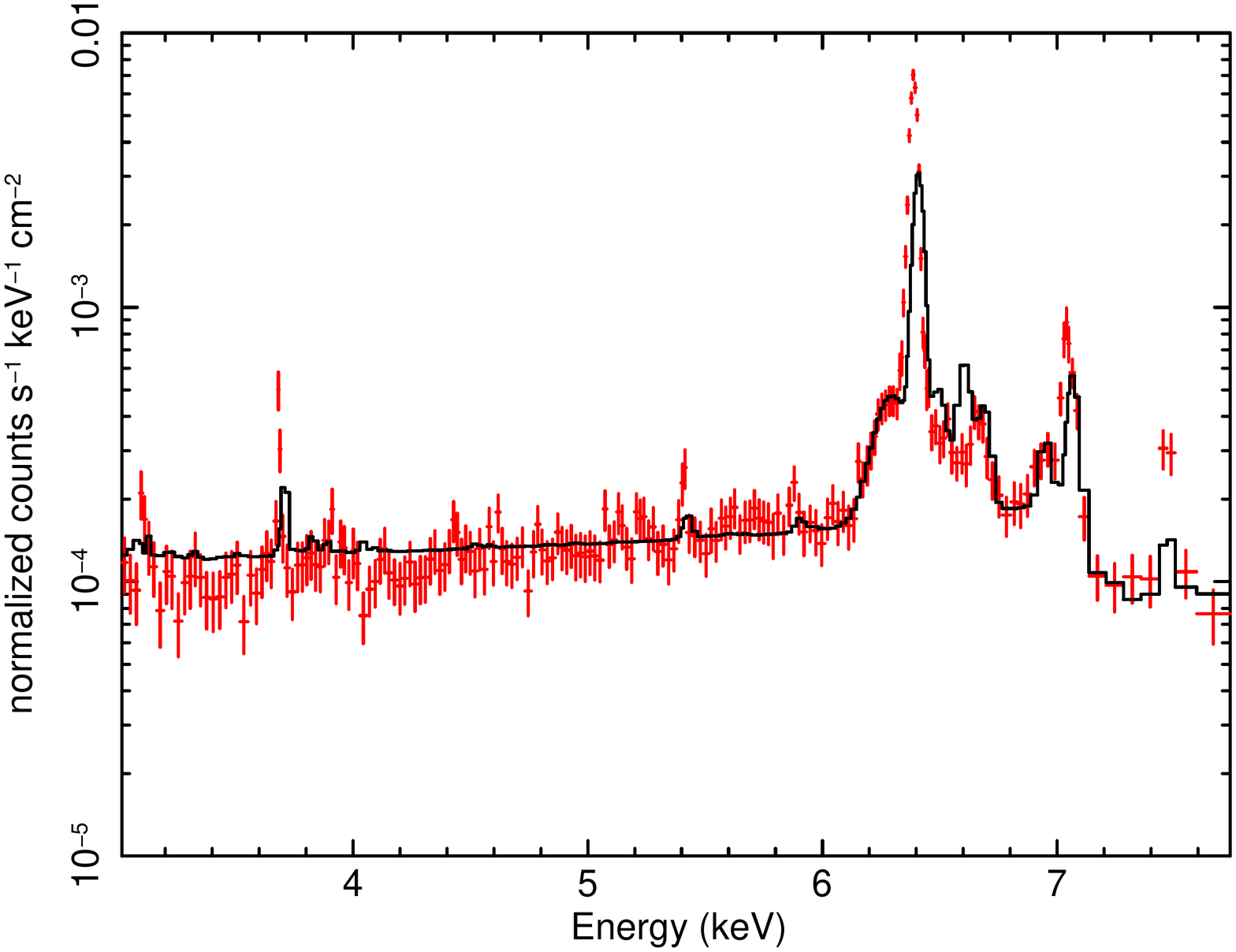}
\caption{
The spectrum of the Circinus galaxy
observed with  \textit{Chandra}/HETG (red crosses). It is 
folded with the energy responses but is corrected for effective area.
The solid curve represents the best-fit model based on our \textsf{Cloudy}
simulations. }
\label{figA-spec}
\end{figure}

We also compare our simulated spectra with the observed data of the
Circinus galaxy, to which the AGN parameters in \citet{Wada2016} are tuned (Section~\ref{sec2.1}).
We utilize the High-Energy Transmission Grating
\citep[HETG:][]{Canizares2005} on Chandra \citep{Weisskopf2002} spectrum as used in
\citet{Uematsu2021}. They analyzed the data from all 13 observations
(observation ID: 374, 62877, 4770, 4771, 10226, 10223, 10832, 10833,
10224, 10844, 10225, 10842, 10843), and created the time-averaged
spectrum for a total exposure of 0.62 Ms. We adopt the model in the
Xspec terminology ``\textsf{phabs} $\times$ 
(\textsf{zcutoffpl} $+$ \textsf{atable\{fountain\_S.fits\}} $+$ \textsf{zgauss})''. The
\textsf{phabs} and \textsf{atable\{fountain\_S.fits\}} terms are the same as those of the
model for NGC~4051. 
The \textsf{zcutoffpl} term also corresponds to the \textsf{zcutoffpl2} term of the
model for NGC~4051. 
The Galactic absorption is fixed at $N_\mathrm{H} = 7.02
\times 10^{21}$~cm$^{-2}$. We add \textsf{zgauss} as the Compton
shoulder component of Fe~K$\alpha$ line at 6.4~keV, which cannot
be reproduced in the \textsf{Cloudy} code, where multiple scattering
processes are ignored.

The observed spectrum and the best-fit model folded with the energy
responses of the Circinus galaxy are plotted in
Figure~\ref{figA-spec}.  We obtain the best-fit inclination angle of
$\sim$87$^\circ$, which is consistent with the previous 
studies, in which the infrared SED, molecular lines in radio, and X-ray continuum spectra are compared with observed ones
\citep{Wada2016,Izumi2018,Uzuo2021,Buchner2021}.  Our model gives a reasonably good
description of observed spectrum except the intensities of the emission
lines. This indicates that additional contributions are needed, as in
the case of NGC~4051. One candidate would be the disk located inside
the torus region, and/or outflows launched from it. This may be in
line with the argument by \citet{Buchner2021} that additional matter inside
$r \sim 0.1$~pc is necessary to account for the shape of the broadband X-ray
spectrum of the Circinus galaxy.

\bibliographystyle{aasjournal}
\bibliography{reference}

\end{document}